\documentstyle[aps,pre,multicol,epsf]{revtex}

\newcommand{\ltue}{``}
\newcommand{\rtue}{''}
\newcommand{\beq}{\begin{equation}}
\newcommand{\eeq}{\end{equation}}
\begin{document}
\title{Collision-induced friction in the motion of a single
  particle on a bumpy inclined line}
\author{S.~Dippel$^1$, G. G. Batrouni$^{1,2}$, D.~E.~Wolf$^1$}

\address{$^1$
H\"ochstleistungsrechenzentrum, 
Forschungszentrum J\"ulich, 
D-52425 J\"ulich, Germany}

\address{$^2$ GMCM, URA CNRS 804, Universit\'e de Rennes 1, 
35042 Rennes Cedex, France}

\maketitle
\begin{abstract}
By means of Molecular Dynamics simulations, we investigate the
elementary process of avalanches and size segregation by 
surface flow in 2 dimensions: a single ball confined to moving along
an inclined line consisting of  balls. The global characteristics of
the motion depend strongly on the 
size of the moving ball relative to the size of the balls on the line,
as well as the distribution of the balls on the line. We find that
in the steady state the friction force acting on the ball is
independent of material properties like 
the Coulomb friction coefficient and the coefficient of
restitution. Contrary to previous notions about the details of the
motion, we find that it is very regular and consists of
many small bounces on each ball on the line. As a result of this
regularity, 
introducing a random spacing between the balls on the line has 
mainly the same influence as a regular spacing of 
adequate length. 
The insensitivity of the steady state velocity to material properties
and to the detailed arrangement of the balls on the line allows for 
an analytical estimation of the mean velocity which fits
the simulation results very well. We find that results from the
2D case can probably not be transferred to
the 3D case of a ball moving on a rough inclined plane as
easily as has been suggested previously.  
\vspace{.25cm}\\
{\small HLRZ Preprint 14/96, submitted to Phys.~Rev.~E, March 1996,
  revised July 1996}
\end{abstract}
\bigskip
 
\begin{multicols}{2}
\section{Introduction}
The flow of granular materials has been studied extensively both
experimentally and theoretically, due to its ubiquity in nature and
its industrial importance. Nevertheless, many properties of granular
flow 
are still poorly understood. Here, we deal with a special case
of granular flow, namely that along an inclined rough
surface. Examples are flow in inclined chutes\cite{dra90} and all
kinds of avalanche processes, such as rockslides \cite{sav89}, which
involve segregation processes (known as \ltue inverse grading\rtue to
geologists), an important phenomenon often encountered in granular
materials \cite{sav93}. In rotating drums, avalanche processes are the
motor of size segregation \cite{can95a,cle95,zik94,bau95}. 
Segregation in surface flow is strongly related to the question of
stability of granular flow on rough surfaces, {\em i.e.~}to the
determination of the limiting conditions for the existence of a steady
state where the flow is neither stopped nor accelerated.  
The rougher the surface encountered by 
the moving particles, the slower they flow and may even come to a 
stop. Thus, the larger the flowing particles are, compared to the
roughness or bumpyness of the surface on which they flow, the farther
they should travel and accumulate (in the case of flow on a sand heap)
at the bottom of the slope. 

The threshold for the onset of granular shear flows as well as the
limiting conditions for which a stable flow ({\em i.e.~}a steady 
state) still exists are insufficiently understood and hard
to determine in realistic situations \cite{cam90}. In particular the
hysteretic properties of 
granular materials, manifesting themselves for example in the
difference between the static and dynamical angle of repose of
sandpiles \cite{fau95}, complicate the situation. Though the static
threshold for the onset of motion on a rough plane ({\em i.e.~}the
tilt of the surface large enough to set a resting mass of granular
material in motion) has been quite thoroughly investigated in
experiments and computer simulations
\cite{pou95b,rig94a}, the dynamical case, {\em i.e.~}the stability 
of the flow, is still poorly understood. 

In order to investigate the
dynamical situation, we follow the course taken up by Riguidel et
al. \cite{rig94a,rig94b,ris94b} in their experimental studies. 
We consider the elementary process of an avalanche: a
single ball of radius $R$ moving down an inclined plane onto which
other balls of radius $r$ are glued. This fixes the roughness of the
surface. Such a system has been investigated experimentally and
numerically in 3D ({\em i.e.~}for a ball moving on a plane)
\cite{rig94a,rig94b,agu95} and in 2D (for a ball moving on a line of
balls) \cite{ris94b,jan92}. The simulations 
in 3D were restricted to the determination of the static angle
of stability for a ball resting on the plane, whereas all experiments
and the simulations on the line  also dealt with the dynamical
situation of a ball that starts on the plane with some initial
velocity, which is the situation we are interested in. In all cases,
three types of motion could be observed, 
depending on the size ratio $\Phi=R/r$ and on the inclination angle
$\theta$ of the plane. These results could be summed up in a
\ltue phase diagram\rtue (typical for both plane and line),
for which we will later give an example from
our simulations. The three types of motion observable in
experiments and simulations are characterized in the following
way. In regime A, the ball gets trapped on the plane, independently of
the initial velocity with which it is launched onto it. In regime B,
it reaches a constant average velocity $\bar{v}$ in the direction
along the plane, in regime C it accelerates
throughout the whole length (2 m in the experiments) of the plane,
accompanied by visible jumps. In the constant velocity regime, 
the mean velocity $\bar{v}$ was found to be proportional to
$\Phi^\alpha\sin\theta$ in 3D, with an exponent $\alpha\approx 1.3$ 
\cite{rig94a,rig94b,agu95}, whereas in 2D $\bar{v}^2$ depended
linearly 
on $\sin\theta$ \cite{jan92,rigdiss}. In 2D, no simple power law
could be found for the $\Phi$-dependence of the mean velocity.
The relation $\bar{v}^2\sim\sin\theta$ was already derived using
very general assumptions by Bagnold for flow of many particles on an
inclined plane \cite{bag54}, but no assumption about the
dimensionality of the system was made there. 

An obvious difference between
the 2D and 3D case is the fact that in 3D, the particle moving down
the plane will be deflected in the direction perpendicular to the
plane inclination, either by rolling down the crooked valleys formed
by the balls on the plane, or by obliquely impacting a sphere on the
plane. We will come back to the importance of this possibility, which
is absent from the simulations we present here, after having discussed
the details of the motion of the ball moving down the line. Recently,
a 2D stochastic model has been proposed which reproduces the angle 
dependence of the velocity in 3D, but gives a different exponent
$\alpha$ \cite{ggb}. 

Here, we discuss the mechanism by which the ball
keeps a constant velocity on the inclined line and how, from the
understanding of 
this mechanism, the transition to the stopping and accelerating regime
can be explained. We will show that the 2D case is {\em generically}
different from the 3D case, i.e.~that the effect of disorder on the
plane cannot be modeled by disorder on a line. Possible reasons for
the success and the 
adequateness of the 2D stochastic model of ref. \cite{ggb} in
describing the 3D case are discussed.  

The outline of the paper is as follows. After presenting our
simulation method in section II, we will first show in section III
that the simulations reproduce  
experimentally observed macroscopic behaviour. We then proceed to a
detailed analysis of the microscopic properties of the motion and give
an explanation of the mechanism stabilizing the motion of the ball in
the 2-dimensional case. The influence of material properties on the
motion is investigated. On the basis of our simulation results, we
present in section IV a simplified model for the motion of the ball
which allows the 
analytical derivation of the mean velocity in the 2-dimensional
case, as well as the determination of the boundary between regions A 
and B in the phase diagram. The results of this model agree with the simulation
results. Possible reasons for the difference between the 2-dimensional
and the 3-dimensional system are discussed. 

\section{Simulation Method}

We model the 2D case by the Molecular Dynamics (MD) technique
\cite{all87}, which was first introduced to the simulation of
granular materials by Cundall and Strack \cite{cun79}. The MD
technique consists of time-integrating Newton's equations of motion
for a system of grains starting from a given initial
configuration. Since our simulations are 2-dimensional, the grains
have only three degrees of freedom, two translational, one
rotational. Two grains of radii $R_i$, positions $\vec{r}_i$, 
velocities $\vec{v}_i$, and angular velocities $\omega_i$
($i = 1, 2$), are in contact when their
(virtual) overlap $\xi = \mbox{max} (0, \,
R_1 + R_2 - \left| \vec{r_2} - \vec{r_1} \right| )$ is larger than
zero (\ltue soft\rtue~grains).  
Two unit vectors $\vec{n}$ and $\vec{s}$ are used to decompose
the forces and velocities into normal and shear components:
\begin{eqnarray}
\vec{n} & = & \frac{\vec{r}_2 -\vec{r}_1}{|\vec{r}_2 -\vec{r}_1|} \\
\vec{s} & = & \left( \left(\vec{n}\right)_y, -\left(\vec{n}\right)_x \right).
\end{eqnarray}
The forces between grains are then given by
\beq 
\vec{F}_{ij} = F_n\vec{n}+F_s\vec{s},
\eeq
 where
\beq
F_n = - k_n \xi - \gamma_n \dot{\xi},
\label{fnorm}
\eeq
\beq
F_s = - \min(|\gamma_s v_s|,|\mu F_n|) \cdot \mbox{sign}(v_s).
\label{ftang}
\eeq
Here $\mu$ denotes the Coulomb friction coefficient. 
The relative normal velocity $v_n$ and the relative shear
velocity $v_s$ ({\em i.e.~} the relative velocity of the surfaces at
the point of contact) are defined as
\begin{eqnarray}
v_n & = & \left(\vec{v}_2 - \vec{v}_1\right) \cdot \vec{n} \\
v_s & = & \left(\vec{v}_2 - \vec{v}_1\right) \cdot \vec{s}
                     + \omega_1 R_1 + \omega_2 R_2.
\end{eqnarray}

A number of different force laws is commonly used in MD simulations of
granular materials; a discussion of their properties can be found in
\cite{schae96}. Our choice of $F_n$ corresponds to a simple linear
spring-dashpot, the tangential force $F_s$ is the Coulomb friction law
for sliding friction, which was regularized for small $v_s$ to avoid
the discontinuity of the Coulomb law at $v_s=0$. The tangential
damping constant 
$\gamma_s$ should have a sufficiently high value such that the case
$F_s= - \gamma_s v_s$ occurs only for very small $v_s$. Only then the
interpretation of $F_s$ a mere regularization of Coulomb friction
holds.
This force law has the advantage of holding equally well in 
free impacts of spheres \cite{schae96,foe94} and in long lasting
contacts \cite{rad96} if the particles can be considered as rough hard
spheres ({\em i.e.~}if tangential elasticity can be neglected). It
also gives a velocity dependent
coefficient of tangential restitution, a feature 
found to be important in a stochastic model of the
situation \cite{ggb,proc}. 

Throughout our simulations we used the parameters $k_n=2\cdot 10^6$
N/m, $\gamma_s=100$ kg/s, $\mu=0.13$, $r=5$ mm and
$M=\frac{4}{3}\pi R^3\rho$ for the mass of the rolling ball with $\rho
= 7.8$ ${\rm g}/{\rm cm}^3$. The values of these parameters were
chosen to match the steel balls used in \cite{ris94b,rigdiss}; the choice
of $k_n$ leads to a collision time of the order of $10^{-5}$ s, which is
a typical value for steel balls of this size. The value of $\gamma_s$ is high
enough such that $F_s$ is reasonably close to the exact Coulomb friction
law in the sense explained above. The damping $\gamma_n$
is determined by fixing the normal coefficient of restitution
$e_n=-v^f_n/v^i_n$, defined by the ratio of final and initial normal
velocities. Unless stated otherwise, the results
presented in the next section were obtained using $e_n=0.7$ and the
above mentioned parameters. The influence of the material parameters
$e_n$ and $\mu$ on the behaviour of the system will be discussed
later. The only external force acting on the ball is gravity, the
gravitational acceleration in the $x$ ($y$) direction is given by
$g\sin\theta$ ($-g\cos\theta$). The integration method we employ is a
constant-timestep fifth order predictor-corrector \cite{all87}.

Fig.~1 shows a
schematic drawing of the ball on the line. The spacing between two
balls fixed on the line is $2 \epsilon r$, where $\epsilon$ is a number
which in the 
disordered case is chosen uniformly distributed in the interval 
$[0,\epsilon_{\mbox{\tiny max}}]$. The {\em impact angle} $\gamma$ is
defined as the angle enclosed by the line joining the centers of the
impacting balls and the normal to the plane. It is taken
to be negative when the ball collides with the uphill 
facing side of a ball on the plane, positive on the downhill side. 

\begin{figure}[tbh]
\centerline{
\epsfysize 6 cm
\epsfbox {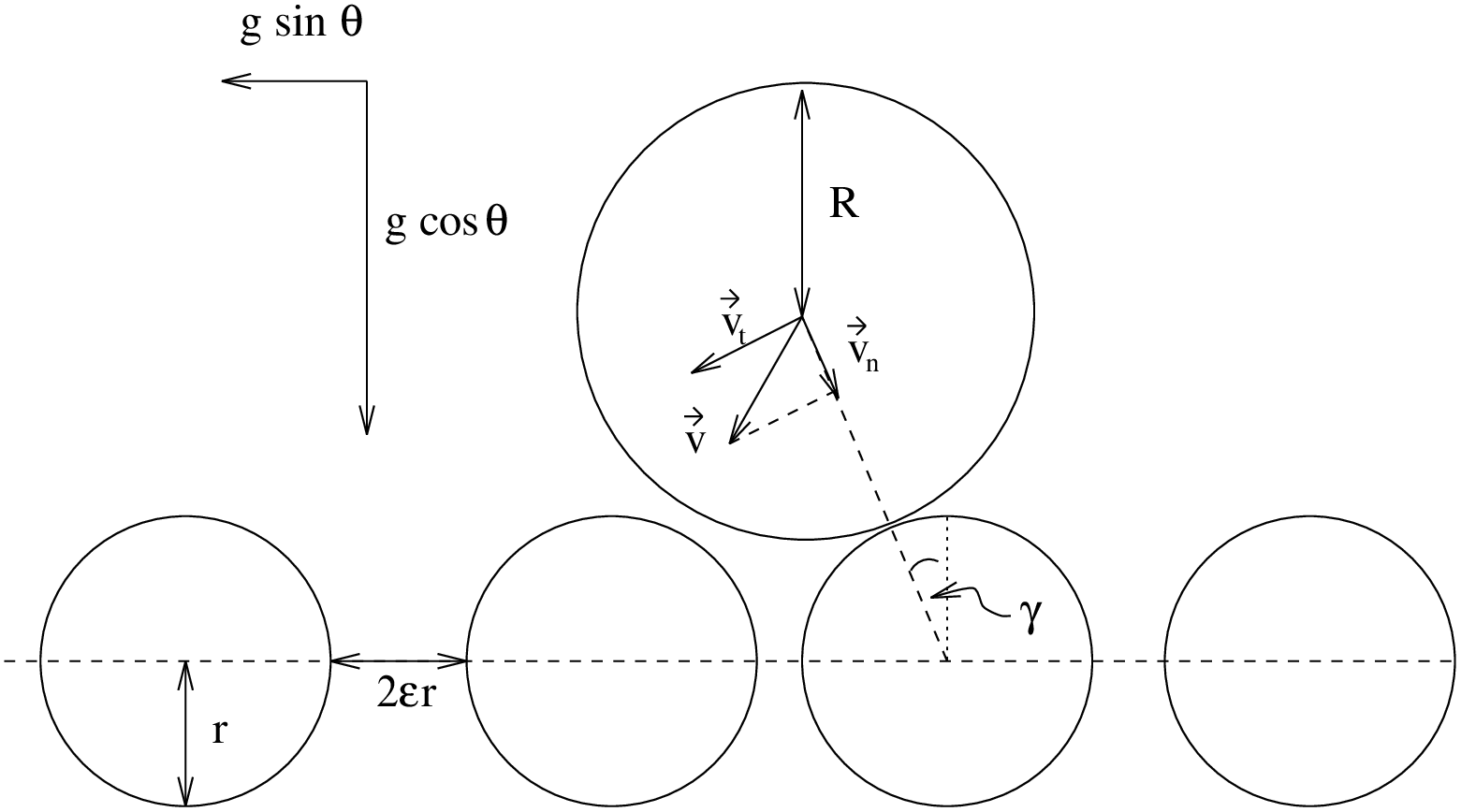}}
{\small FIG.~1. Schematic drawing of the ball on the line.}
\end{figure}

\section{Simulation results}
\subsection{Global characteristics of the motion}

In all simulations presented here, the ball was launched onto the line
with a rather high velocity $v_x$ in the $x$--direction and quite low
$v_y$. If inclination angle $\theta$ and size ratio $\Phi=R/r$ are in
a suitable range, the moving ball very quickly, usually after passing
only a few balls on the line, reaches a steady state with
well-defined mean velocities in the $x$-- and $y$--directions. Clearly,
the average over $v_y$ is zero in our problem, so the 
only interesting {\em mean} velocity is the average over $v_x$, which
we denote by $\bar{v}$. To obtain this
mean velocity, we first average over a certain number of timesteps (usually
500). This value is large enough to average out the comparably large
fluctuations occurring during collisions, while it is still so small
that the ball moves only a very short distance (much less than the
radius of balls on the line)
during this time. This averaged value clearly
still gives a fluctuating velocity, but the fluctuations 
are very small, and $\bar{v}$, the mean value of these averaged
velocities, is well-defined. 

We investigated the motion of the ball both on lines with equally
spaced balls and on lines with randomly spaced balls. We found that
the essential features of the motion are alike in both cases. 
Fig.~2 shows the velocity of the rolling ball for various $\Phi$ and
equal spacing of balls on the line with $\epsilon=0$ as obtained from
simulations using the above mentioned parameters. 
For the case $\epsilon=0$, experimental data is available (though only
for size ratios $\Phi \le 2.0$) \cite{ris94b,rigdiss}. Our stable
mean velocities are of the same order of magnitude, but in the
simulations the range of inclination angles for which $\bar{v}$ is
well--defined, {\em i.e.~}a steady state is reached, 
is a bit narrower than in the experiments. This might be due
to the difference in 
the experimental setup. There, a ball moved down a line of balls
sitting in a V-shaped groove, and it is to be expected that contact
with the groove walls influenced the motion as an additional source of
dissipation due to friction and collisions with the walls, but it is
unclear how strong this influence was. For a direct comparison
of simulation and experiment, it would be desirable that a ``more''
2--dimensional experiment be done, like for example a 
ball moving down a row of cylinders, to rule out these boundary
effects. 

\begin{figure}[tbh]
\centerline{
\epsfysize 5 cm
\epsfbox {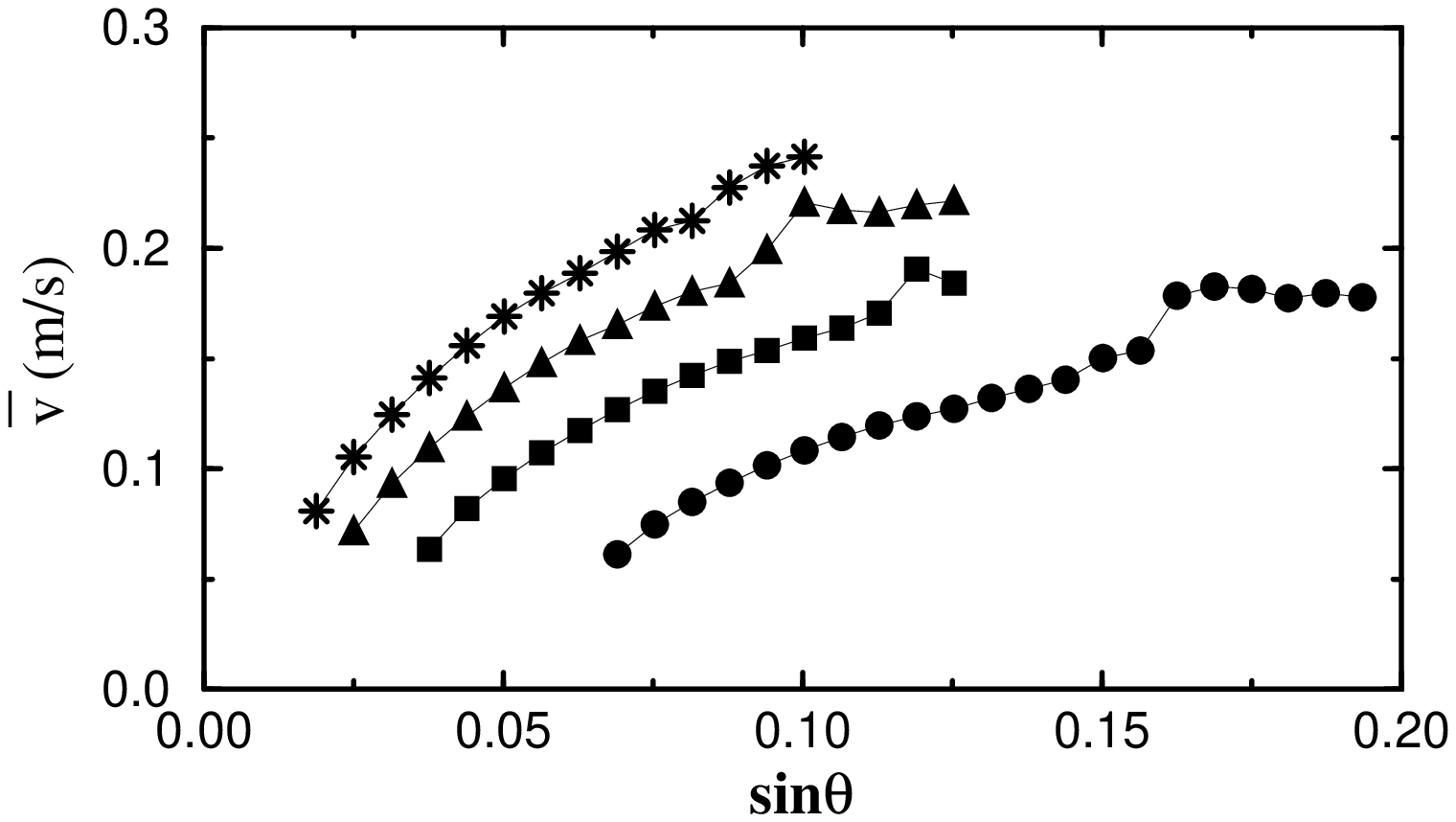}}
{\small FIG.~2. Dependence of the mean velocity $\bar{v}$ on the
  inclination of the line for various size ratios $\Phi$:
  $\Phi=1.5$ (circles), $\Phi=2.0$ (squares), $\Phi=2.5$
  (triangles), $\Phi=3.0$ (stars).}
\end{figure}

For angles lower than the one for which the smallest steady state
velocity is reached for a given $\Phi$, the ball loses all the initial
velocity it had 
and very quickly stops, usually after passing only very few 
balls on the line (region A of the phase diagram). We will denote this
minimum angle, for which a steady state with $\bar{v}\neq 0$ still
exists, by $\theta_{\mbox{\tiny AB}}(\Phi)$. This defines the phase
boundary between region A and B.

All velocity curves in Fig.~2 exhibit a sudden increase of $\bar{v}$
to a value where it remains roughly constant. This sudden increase
has also been observed in experiments \cite{ris94b}. For angles
smaller than 
the one where this happens, the velocity of the ball
shows only very small fluctuations and the steady state velocity
$\bar{v}$ does 
not depend on the initial velocity of the ball. At the inclination
angle where the velocity suddenly shoots up, the
behaviour changes qualitatively. The fluctuations of $v_x$ increase
significantly, and the behaviour of the ball now
depends on the initial velocity. Depending on the starting velocity,
the ball can accelerate (usually if the starting velocity is larger
than the stable mean velocity) and start to jump visibly, or it can
reach a constant velocity (if released with an initial velocity
smaller than the stable $\bar{v}$ in this region). However, even when
the ball does not accelerate, in this region 
$\bar{v}$ can depend somewhat on the initial velocity.  
We thus denote the angle at which the sudden increase takes place by 
$\theta_{\mbox{\tiny BC}}(\Phi)$, since it defines the upper boundary
of the constant velocity region B independently of the initial
velocity of the ball. 

If $\theta$ increases even more, the ball accelerates and starts to
jump 
significantly (the length and height of the jumps reaches a few ball
diameters). We did not investigate this jumping motion any further,
for the following two reasons. In the simulations
the ball accelerated up to 50 m/s and more, which is a velocity the
ball would never reach in experiments due to air resistance. Thus, the
question of whether the ball in the jumping regime can reach a steady
state only due to collisions with the plane will not be discussed
here, even though it is of theoretical interest. The second reason is 
that for
grazing impacts at high velocities, an artefact of constant timestep
algorithms, the so-called brake failure effect \cite{schae95} may set
in, which leads to anomalous dissipation of energy. Its onset can be
shifted to higher velocities by increasing the  
spring constant $k_n$, {\em i.e.~}the stiffness of the balls. This,
however, decreases the timestep, thus increasing the simulation time
tremendously. The constant timestep algorithm then becomes a very
ineffective way of simulating the motion, since the ball spends most
of its time in free flight, where a timestep small enough to integrate
collisions correctly and to avoid spurious effects is essentially a
waste of computation time. In this regime, event driven (ED)
simulations would be more appropriate, which is why we will not
discuss the motion in the ``high bounce'' regime. One might
argue that, since the ball moves down the line in a series of bounces,
an ED algorithm might in any case be a more efficient and appropriate
way to simulate the motion. As we will see later, this is not the
case, as both short and long-lasting contacts occur, the latter of
which are not treatable by an ED algorithm in a straightforward way 
\cite{mcn92}. 

The velocity curves obtained from experiments as well as from
simulations suggest the functional form
\begin{equation}
\bar{v}=v_0(\Phi)+f(\Phi)\sqrt{\sin\theta - \sin\theta_{\mbox{\tiny
      AB}}(\Phi)} 
\label{eq1}
\end{equation}
for a fixed value of $r$ in region B. 
All curves start at a certain offset
velocity $v_0$, which seems to depend slightly on $\Phi$. $f(\Phi)$
denotes a (still unknown) scaling function, which, unlike in 3D, does
not seem to be a simple power law.
We thus plot $(\bar{v}-v_0(\Phi))^2$ in Fig.~3, where $v_0(\Phi)$ is
obtained by 
fitting a square root to $\bar{v}(\sin\theta)$. $v_0(\Phi)$ typically
is of the order of 4 cm/s. The error bars give the variance of the
averaged velocities, averaged (as the plotted value of $\bar{v}$
itself) over a number of simulation runs with different starting
velocities. Fig.~3a shows
velocities in the case $\epsilon=0$ for various size ratios,
Fig.~3b illustrates the influence of disorder on the velocity of a
ball of size ratio $\Phi=2.25$.

\begin{figure}[tbh]
\centerline{
\epsfysize 5 cm
\epsfbox {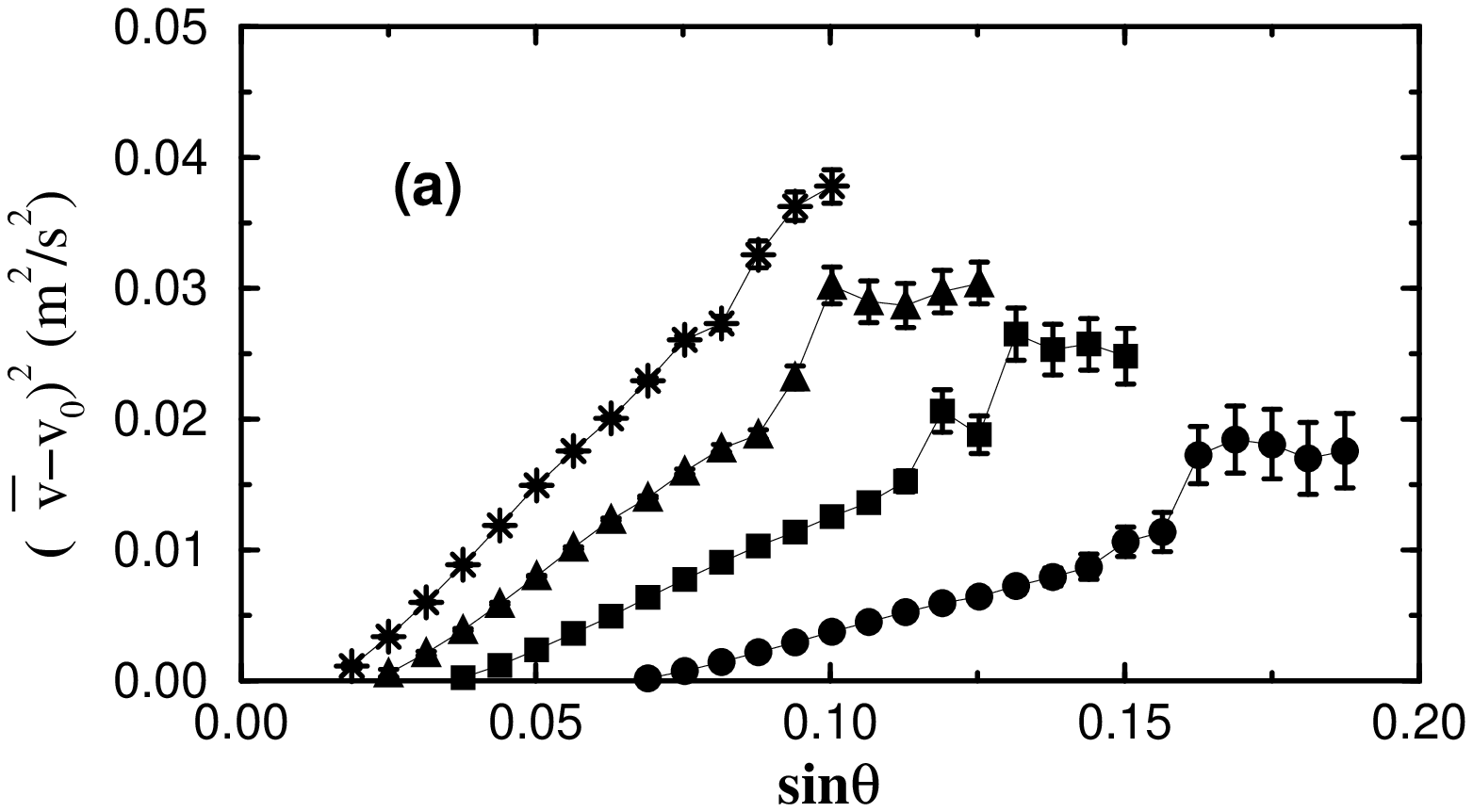}}
\centerline{
\epsfysize 5 cm
\epsfbox {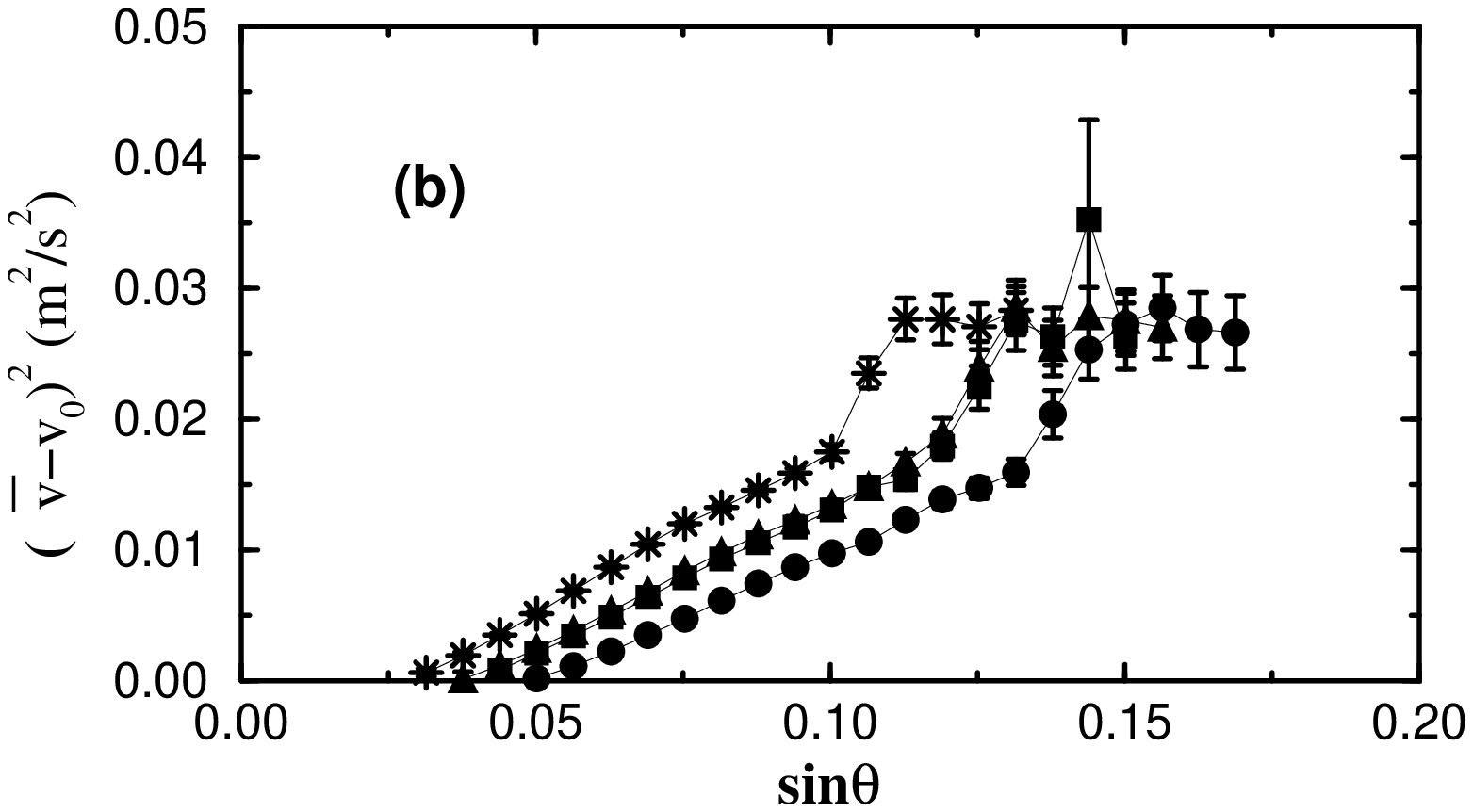}}
{\small FIG.~3. Dependence of the mean velocity $\bar{v}$ on the
  inclination of the line (a) for the same parameters as Fig.~2; (b)
  for various 
  spacings of the balls on the line ($\Phi=2.25$): balls equally
  spaced with $\epsilon=0$ (stars), $\epsilon=0.1$
  (triangles), $\epsilon=0.2$ (circles); balls disordered
  with $\epsilon_{\mbox{\tiny max}}=0.2$ (squares). }
\end{figure}

Fig.~3b demonstrates how the velocity of the ball
is affected by the introduction of disorder to the line. It shows that
$\bar{v}$ on the disordered line with $\epsilon_{\mbox{\tiny
    max}}=0.2$ can be approximated by the 
velocity on a line with equally spaced balls and a spacing
$\epsilon=0.1$, corresponding to the mean value of the disordered
case. It also shows that $\theta_{\mbox{\tiny BC}}(\Phi)$ depends on both $\Phi$ and
the  
arrangement of the balls on the line, whereas the maximum $\bar{v}$
only seems to depend on $\Phi$.

In Fig.~4 we plot the corresponding phase diagram 
for three cases: two cases for a line with balls
equally spaced, with $\epsilon=0$ and $\epsilon=0.2$, the third
for a disordered line with $\epsilon_{\mbox{\tiny
    max}}=0.2$. The lines denote the phase boundaries given by the
angles $\theta_{\mbox{\tiny AB}}(\Phi)$ and $\theta_{\mbox{\tiny
    BC}}(\Phi)$ defined previously. Obviously, the introduction of
disorder has the same 
effect on the phase boundary AB as the introduction of an equal
spacing of balls with $\epsilon=\epsilon_{\mbox{\tiny max}}$. This
is understandable; the stopping of a ball should be ruled by the
deepest ``traps'' into which it might fall. The boundary BC
rather seems to be determined by the mean spacing of balls, since in
the disordered case it falls somewhere between the two extreme cases
of an ordered array with $\epsilon=0$ and
$\epsilon=\epsilon_{\mbox{\tiny max}}$. 

\begin{figure}[tbh]
\centerline{
\epsfysize 5 cm
\epsfbox {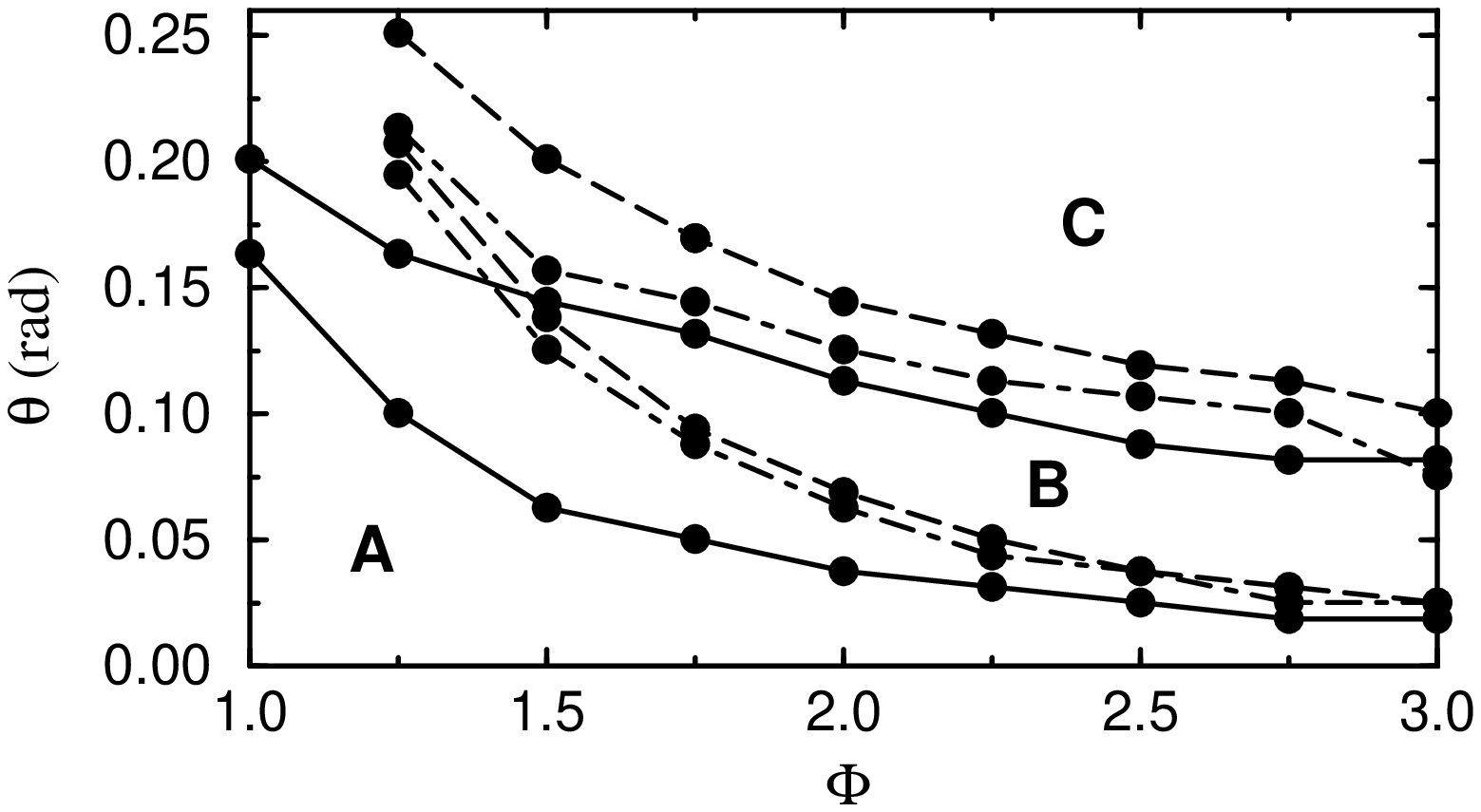}}
{\small FIG.~4. Phase diagram for $e_n=0.7$ and various spacings of the
  balls on the line: balls equally spaced with $\epsilon=0$
  (solid line) and $\epsilon=0.2$ (dashed line); balls disordered
  with $\epsilon_{\mbox{\tiny max}}=0.2$ (dot--dashed line).  }
\end{figure}

\subsection{Detailed dynamics of the motion}

In order to investigate the mechanism by which the ball maintains a
constant velocity in region B, we have to look into the details of the
motion. Certainly, distributions of the impact angle $\gamma$, of
times of flight between impacts or of the impact velocity $\vec{v}^i$
would be of interest. One might ask as well if there are
correlations between impacts at certain angles and the corresponding
impact velocities or between the time of flight after an impact and
the corresponding impact angle. Actually, the latter provides even
more detailed information; for this reason we will first have a look
at these correlations.
We will first discuss the ordered case of a line with no spacing
between the balls, {\em i.e.~}with $\epsilon=0$.

In Fig.~5 we plot the velocity of the ball right 
before an impact as a function of the corresponding impact angle
$\gamma$. Each dot corresponds to a collision. We chose to plot the
normal velocity $v_n$ and the 
tangential {\em translational} velocity $v_t=\vec{v}\cdot\vec{s}$
rather than $v_x$ and $v_y$, as they provide more information on the
mechanisms involved. Since dissipation takes place only through the
normal velocity (dissipation due to $F_s$ is negligible, as we will
see later), the evolution of this quantity may explain the mechanism of
energy loss maintaining the steady state. Since $v_t$ is the velocity
component of the motion of the center of mass of the ball along the
bumps of the surface, it reflects how this bumpyness is felt. It is
obvious that there is 
a strong correlation between 
the impact angles and the corresponding impact velocities. The times
of flight between 
impacts and corresponding previous impact angles are equally strongly
correlated (see Fig.~6). From both Fig.~5 and 6 it is obvious that the
ball moves down the line in a series of bounces. There is even a
certain range of impact angles that are never hit. In addition, we can
extract from Fig.~6 that the typical times of flight between impacts 
correspond to a distance of less than 3 mm, which is smaller than the
radius of the balls constituting the 
line. So the bounces the ball undergoes cannot be very high or far,
and the bouncing ball collides with every ball on the line 
several times.

\begin{figure}[tbh]
\centerline{
\epsfysize 5 cm
\epsfbox {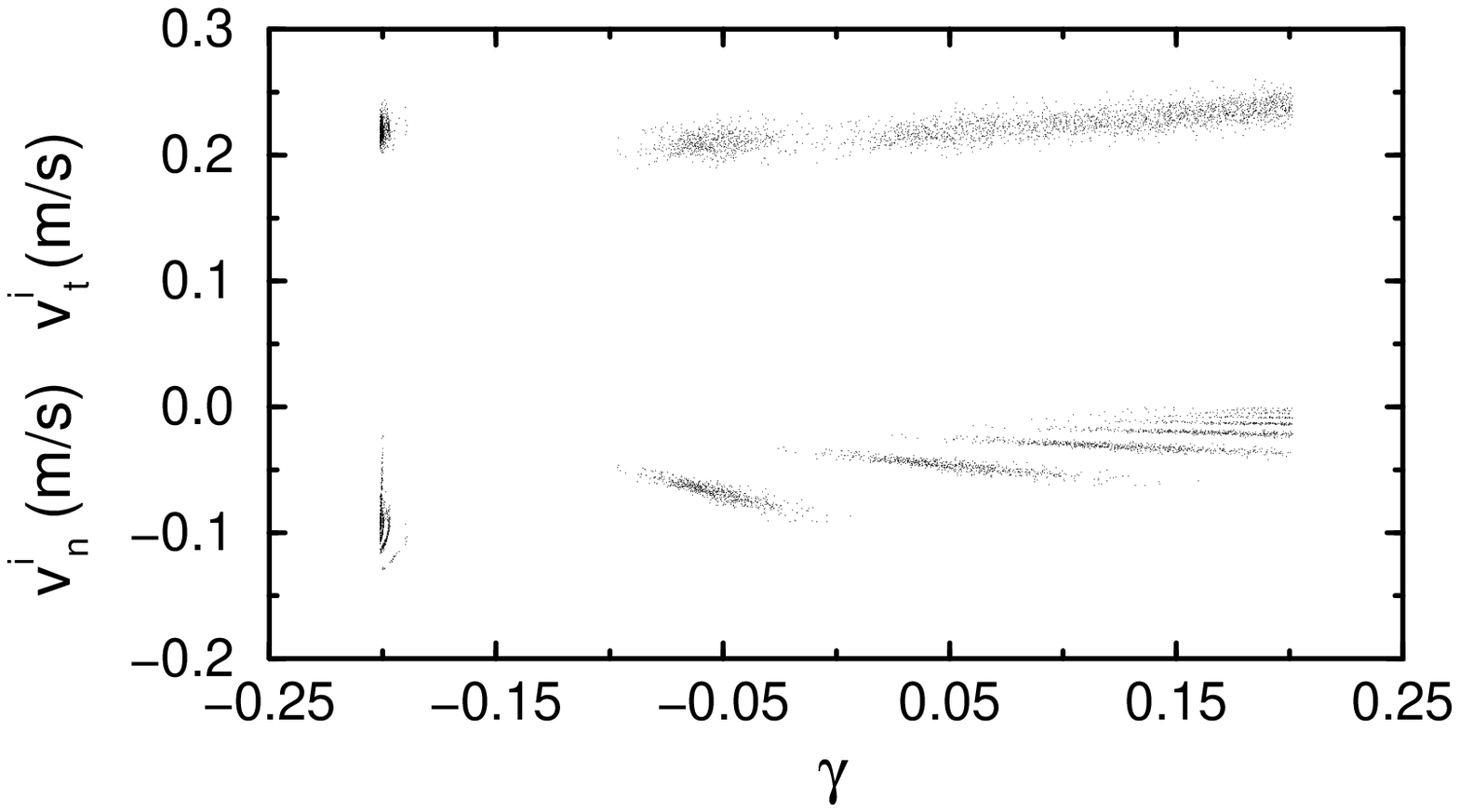}}
{\small FIG.~5. Correlation of impact velocities and impact angles for
  $\Phi=4$, $\sin\theta=0.05$, $\epsilon=0$. The upper points
  correspond to the relative translational tangential velocity
  $v_t=\vec{v}\cdot\vec{s}$, the lower ones to the relative normal
  velocity $v_n$. }
\end{figure}

\begin{figure}[tbh]
\centerline{
\epsfysize 5 cm
\epsfbox {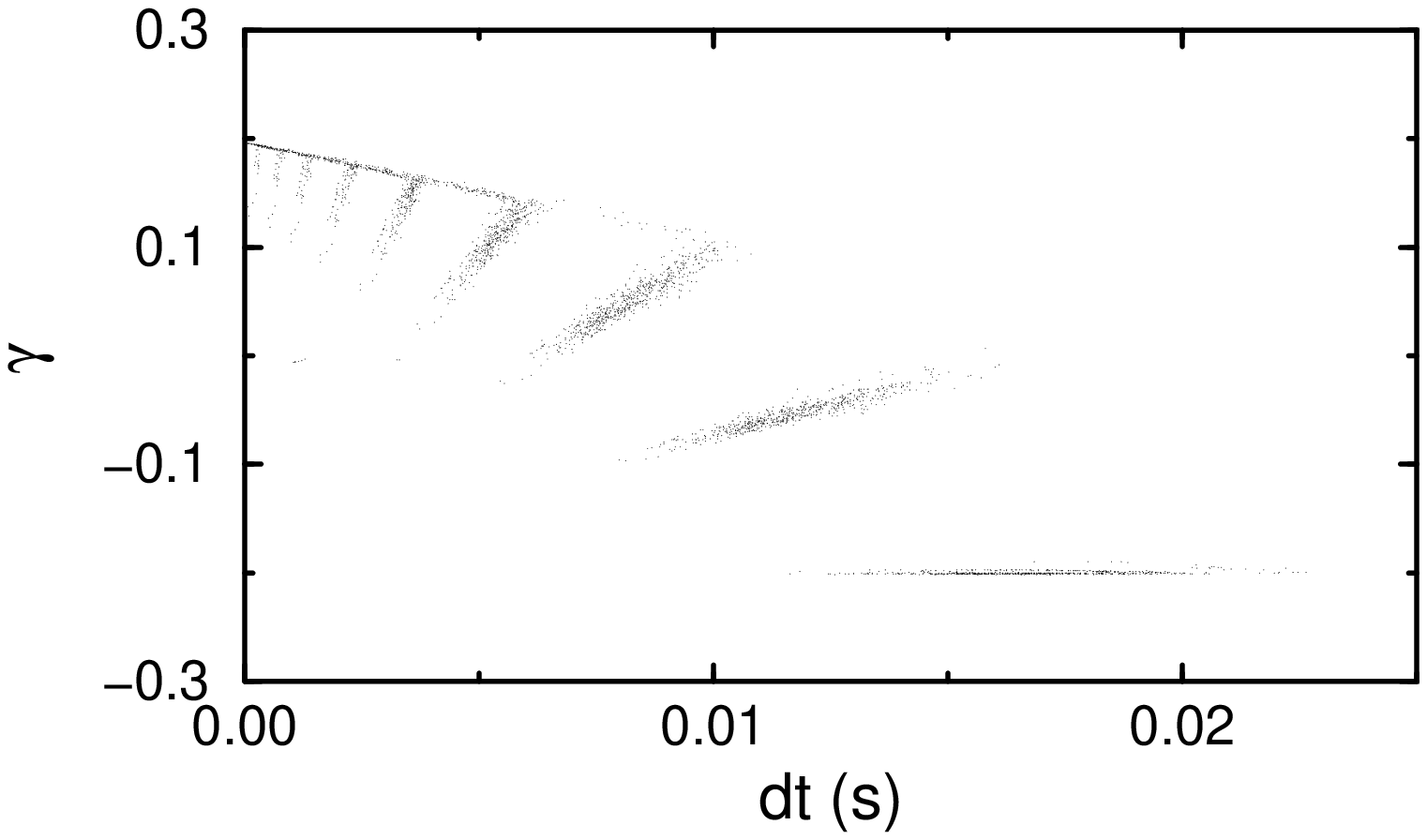}}
{\small FIG.~6.  Correlation of time of flight after an impact at angle
  $\gamma$ and corresponding previous impact angles (parameters as in
  Fig.~5).} 
\end{figure}

How can these results be understood?
From Fig.~5 two important points can be extracted. Firstly we see that
at certain impact angles $\gamma$, the moving ball is 
likely to hit the ball on the line with a well-defined corresponding
normal velocity $v_n$, which leads to an equally well-defined time of
flight (see Fig.~6). Secondly, the normal velocities are largest for
negative impact angles ({\em i.e.~}hitting on the uphill side of a
line ball) and get smaller (and eventually very close to zero) with
increasing impact angles. The second observation helps to explain the
correlations and reveals the reasons for the regularity of the motion.

Consider a ball that has just arrived at the maximum possible
positive impact angle $\gamma_{\mbox{\tiny max}}$ on a certain ball on
the line, say ball number $k$. The value of $\gamma_{\mbox{\tiny
    max}}$ is given by the geometry and defined as
\beq
\gamma_{\mbox{\tiny max}}=\arcsin \frac{1+\epsilon}{1+\Phi}. 
\eeq
From Fig.~5 we see that at $\gamma_{\mbox{\tiny max}}$ the moving ball  
has lost nearly all normal velocity with respect to ball $k$,
{\em i.e.~}it is rolling or sliding down the lower part of the \ltue
downhill side\rtue of the corresponding ball on the line. Even though
the total shear velocity $v_s$ at the point of contact is quite small
(the ball rolls, thus rotational velocity is opposed to the
translational tangential 
velocity and nearly compensates it), the translational tangential
velocity $v_t=\vec{v}\cdot\vec{s}$, which we plot in Fig.~5, is quite
large (close to $\bar{v}$ for larger $\Phi$). Immediately
after having reached $\gamma_{\mbox{\tiny max}}$ on ball $k$, the
moving ball impacts the next ball, $k+1$, at $-\gamma_{\mbox{\tiny
    max}}$, where a large part of this previously tangential velocity
is now normal velocity, as the direction of the vector $\vec{n}$
connecting the centers of the impacting balls with respect to
$\vec{v}$ has changed. The ball thus gets thrown up again, and manages
to reach the downhill side of ball $k+1$ after a few jumps. The
comparably high 
normal velocity of the ball at $-\gamma_{\mbox{\tiny max}}$ is lost in
two ways in crossing a fixed ball: most is lost by dissipation due to
impacts, but 
partly it is converted into tangential velocity by the increasing
obliqueness of successive impacts at positive $\gamma$. In the process
of bouncing over the top of ball $k+1$ on the line, the 
moving ball loses nearly all of its normal velocity, so that it again
reaches $\gamma_{\mbox{\tiny max}}$ with nearly only tangential
velocity. These steps repeat
themselves over and over again while the ball moves down the plane,
thus retaining a constant mean velocity.
Note that strong geometrical constraints prevent the ball from rolling
down the line without ever bouncing. Whenever the moving ball
rolls down the downhill side of a ball $k$, it is thrown onto the
uphill side of ball $k+1$ with considerable normal velocity with
respect to ball $k+1$ as a result of the change in the geometry. To
remain in contact with ball $k+1$, the 
moving ball would to have to lose a very substantial amount of this
normal velocity in a single impact, or it would jump up. Persistent
rolling thus is only possible in 
the case of vanishing normal restitution. Rolling on part of each ball
is possible, however, as we will see in a moment.

In order to describe things more quantitatively, we take a look
at the distribution of impact angles. Fig.~7a gives an example for an
angle of inclination in the middle of region B of the phase diagram
for $\Phi=2.5$. The distribution exhibits clear peaks for $\gamma <
0$. As $\gamma$ increases, the peaks broaden somewhat and approach
each other until they are nearly indistinguishible in the
histogram, although there still is structure in
the correlation plots. For plane inclinations closer to
$\theta_{\mbox{\tiny AB}}$ than $\theta_{\mbox{\tiny BC}}$, the
distribution, like in Fig.~7a, even breaks 
off at a value $\gamma < \gamma_{\mbox{\tiny max}}$, indicating the
start of a long-lasting contact. Here, the ball has lost
so much of its normal velocity in impacts it suffered on crossing over
the top of a  
ball on the line, that it finally starts to roll over part of
this ball. This rolling motion can even start while the moving ball is
still on the uphill facing side of the fixed ball, though this only
takes place very close to $\theta_{\mbox{\tiny AB}}$. All these
impacts {\em end} at $\gamma_{\mbox{\tiny max}}$, which leads to the
very pronounced peak at $-\gamma_{\mbox{\tiny max}}$. 
Integrating the distribution over negative and
positive angles respectively yields the result that there
are actually more impacts for positive than for negative $\gamma$. 

\begin{figure}[tbh]
\centerline{
\epsfysize 10 cm
\epsfbox {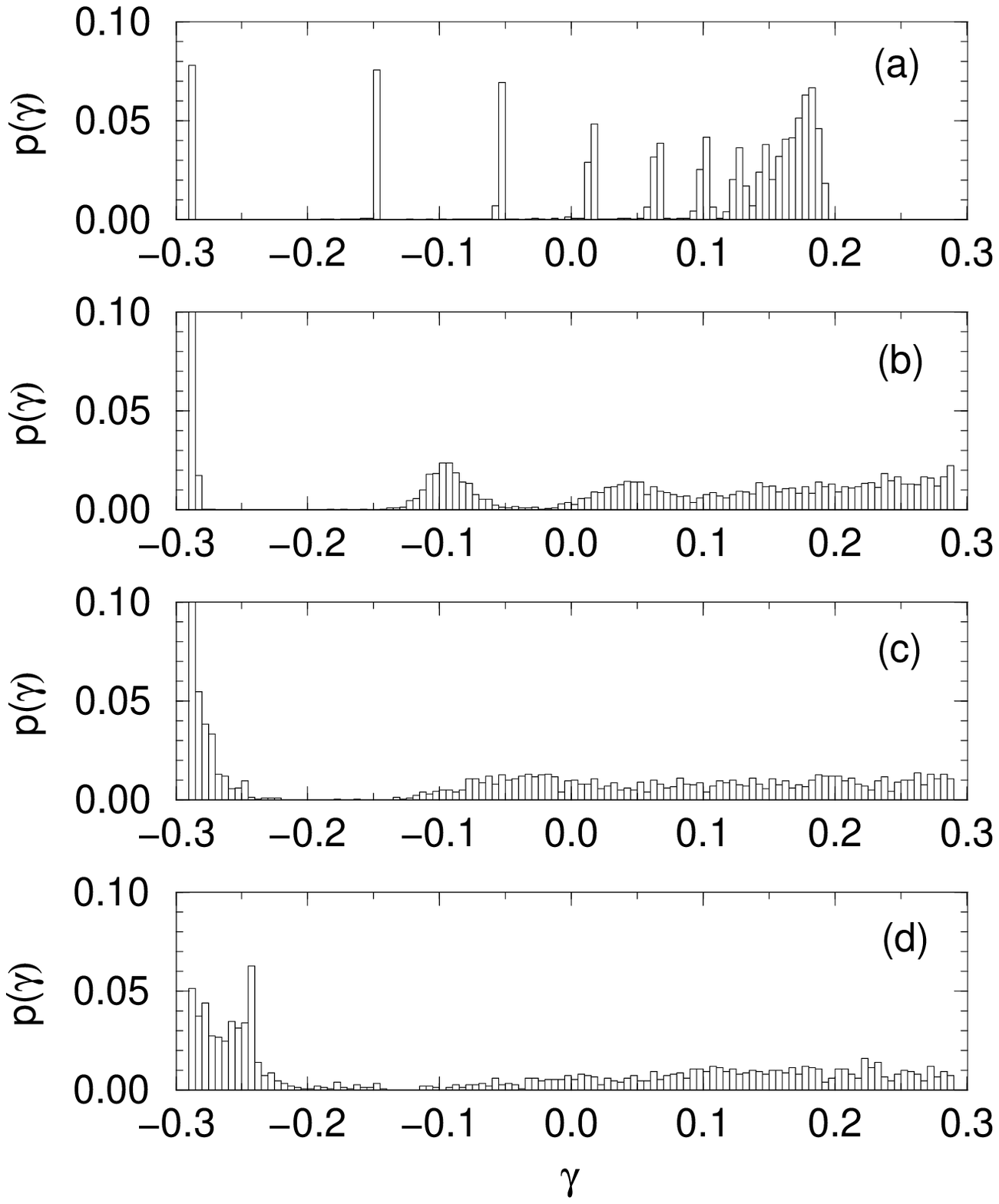}}
{\small FIG.~7. Distribution of impact angles for $\Phi=2.5$,
  $\epsilon=0$ and (a) $\sin\theta=0.057$ (b) $\sin\theta=0.075$ (c)
  $\sin\theta=0.091$ (d) $\sin\theta=0.103$.}
\end{figure}

When $\theta$ is increased, the peaks for negative $\gamma$, except
the one at $-\gamma_{\mbox{\tiny max}}$, move towards zero, finally
disappearing into the continuous distribution for positive 
$\gamma$ (see Fig.~8). At a large enough inclination, they get
visibly broader, showing that 
the velocity at $\gamma_{\mbox{\tiny max}}$ is now no longer as
sharply defined as before, probably due to the fact that not all the
normal velocity can 
be lost on only one ball. Still, the ball does not accelerate as long
as its starting velocity is only a little higher than its stable mean
velocity (this is the region beyond the transition from linear to more
irregular behaviour in Fig.~2).

Fig.~7(b-d) shows how the motion of the ball changes qualitatively at
$\theta_{\mbox{\tiny BC}}$. The $\theta$-values are taken a little below
(b), as close as possible to (c) and a little above
$\theta_{\mbox{\tiny BC}}$ (d). The heretofor 
clearly defined peak at $-\gamma_{\mbox{\tiny max}}$ broadens
considerably and even
seems to develop a small side peak as $\theta$ reaches
$\theta_{\mbox{\tiny BC}}$. The qualitative change of the behaviour of the ball
can also be observed in the velocities.  
In particular in the angle region where the 
velocity curve flattens out again, an intermittent behaviour of the
ball can be observed - it will accelerate a little, even start to jump
a little, but then suddenly get braked again. The mean velocity in
this case is determined by how fast the ball is accelerated and
braked respectively. The same intermittent behaviour of the ball
shortly before passing from motion with a mean constant velocity to 
a jumping regime has been observed in 3D experiments
\cite{rigdiss} and in a stochastic model for the 2D case
\cite{ggb}.

\begin{figure}[tbh]
\centerline{
\epsfysize 5 cm
\epsfbox {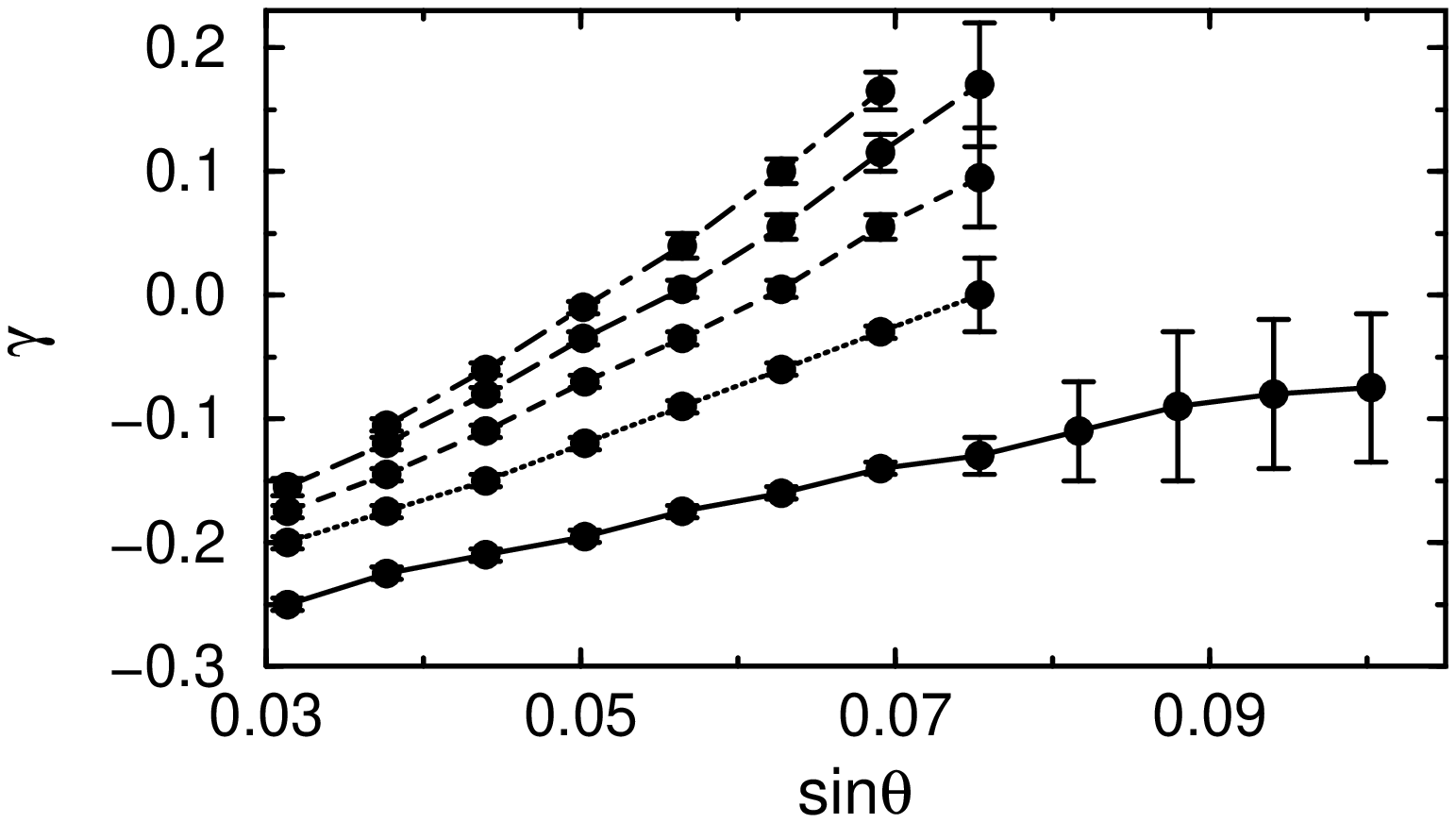}}
{\small FIG.~8. Location of the peaks in the distribution of impact
  angles for $\Phi=2.25$. The error bars denote the width of the
  peak. The solid, dotted, dashed, long-dashed and dot-dashed lines
  denote respectively the 1st, 2nd,..., 5th distinguishable peak,
  excluding the one at $-\gamma_{\mbox{\tiny max}}$.}
\end{figure}

While the intermittent
behaviour marks the transition from region B to region C in the phase
diagram, stopping of the ball takes place when the typical tangential
velocity at $\gamma_{\mbox{\tiny max}}$ does not convert into enough
normal velocity at $-\gamma_{\mbox{\tiny max}}$ to carry the ball over
the top of the
next ball in the line in a few jumps, or at least up to a point where
the remaining tangential velocity suffices to make it roll over the
top of this ball.

In regime B, the motion is not only characterized by typical impact
angles $\gamma$, but also by a very strong correlation of successive
impact angles. In Fig.~9, we plot $\gamma_{\mbox{\tiny n+1}}$, the impact angle for
impact $n+1$ as a
function of the previous impact angle $\gamma_{n}$. Even when
introducing a 
random spacing of balls on the line the strong correlation remains.
This observation leads us to the discussion of the behaviour of the
moving ball on a disordered line. The phase diagram
already suggests that introducing disorder has a similar influence on
the motion of the ball as introduction of a regular spacing. In both
cases, the region of stable motion in the phase diagram 
shifts to larger $\Phi$ and $\theta$. Plotting the distribution of
impact angles for the cases displayed in Fig.~3b shows the effect of
disorder (see Fig.~10). 

\begin{figure}[tbh]
\centerline{
\epsfysize 5 cm
\epsfbox {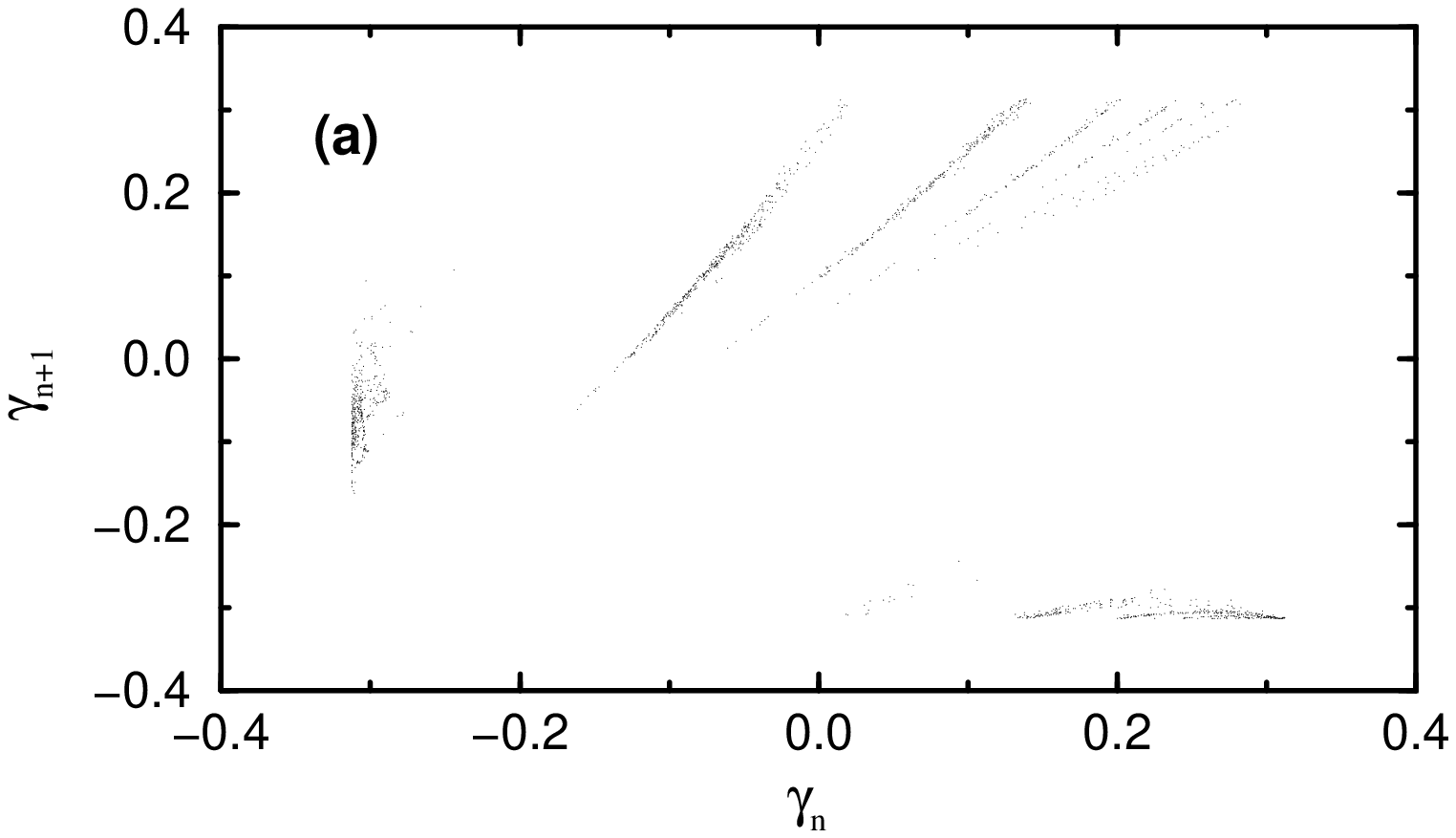}}
\centerline{
\epsfysize 5 cm
\epsfbox {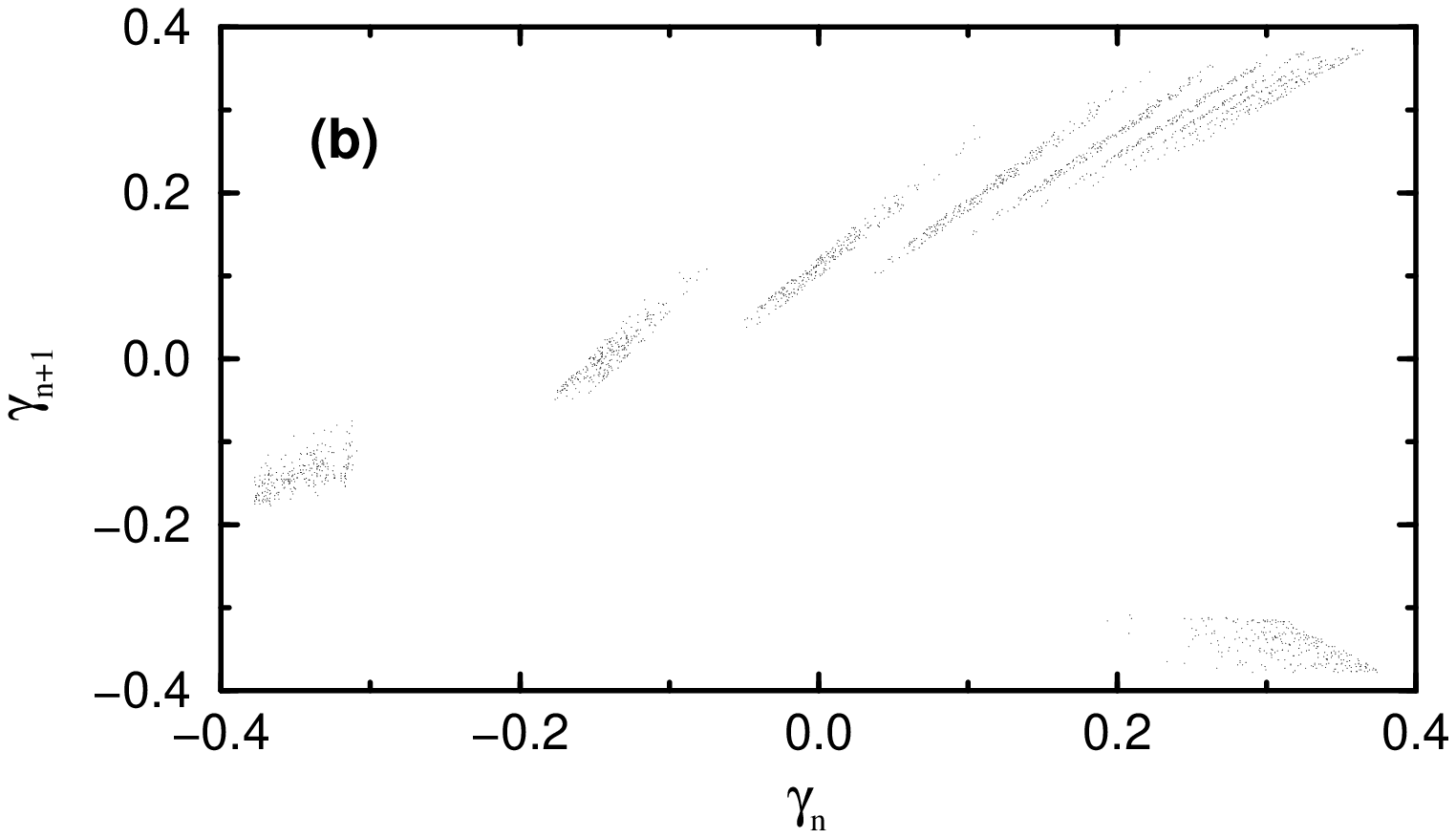}}
{\small FIG.~9. Correlation of successive impact angles for $\Phi=2.25$,
  $\sin\theta=0.094$. (a) Balls on the line regularly spaced
  ($\epsilon=0$), (b) balls on the line disordered with
  $\epsilon_{\mbox{\tiny max}}=0.2$.}
\end{figure}

We checked that this regularity in the case of a disordered line is
not a
finite size effect. The angle distributions do not change when the
size of the system is increased.
From Fig.~10 it can be seen that the distributions for the impact
angles in the disordered case lie between the limiting
cases $\epsilon=0$ and $\epsilon=0.2$ of an ordered line. In addition,
the peaks for $\epsilon=0.1$ correspond to the center of the peaks in
the disordered case. The
reason for the large width of the peaks in the case
$\epsilon=0$ lies in the fact that here the velocity is already
quite high (see Fig.~3b). We interpret the results of the disordered
case in the following way. The motion of the ball is influenced mainly
by two factors - the minimum and maximum spacing of balls on the
line. If the ball is to keep its mean velocity, without being stopped
and without being accelerated, the velocity has to have a value that
enables it to get out of the \ltue deepest valleys\rtue existing
between two balls (given by $\epsilon_{\mbox{\tiny max}}$), but still is low
enough that in all cases (even in those where $\epsilon=0$) most of
the normal velocity is dissipated on crossing over the corresponding
ball so that at $\gamma_{\mbox{\tiny max}}$ of this ball only tangential velocity
is left. The 
variations in this tangential velocity are so small that they only
broaden the peaks for the impact angles, but do not lead to a {\em
  qualitative} change of the motion of the ball.

\begin{figure}[tbh]
\centerline{
\epsfysize 10 cm
\epsfbox {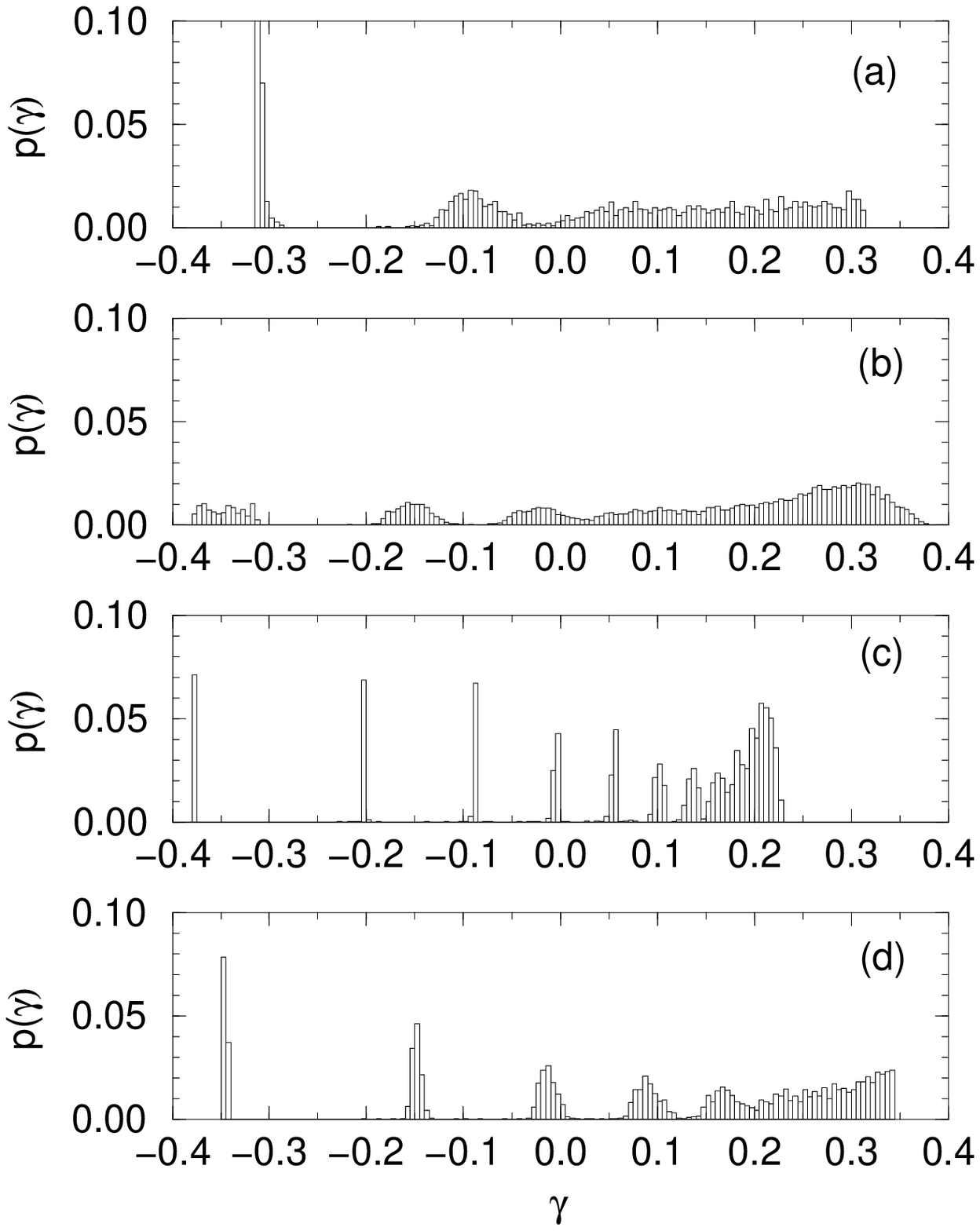}}
{\small FIG.~10. Distribution of impact angles for $\Phi=2.25$,
  $\sin\theta=0.088$ and various spacings of the balls on the
  line. (a) regular spacing, $\epsilon=0$ (b) disordered line,
  $\epsilon_{\mbox{\tiny max}}=0.2$ (c) regular spacing, $\epsilon=0.2$ (d) regular
  spacing, $\epsilon=0.1$. }
\end{figure}

\subsection{Influence of material properties on the motion}

Having uncovered the mechanism by which the ball moves down the line
and keeps a constant average velocity, we now have to ask how much
this mechanism and the global results like $\bar{v}$ are influenced by
material properties. The material properties incorporated in our
simulations are the coefficient of normal restitution $e_n$ and the
friction coefficient $\mu$. We will show in this subsection that their
influence on $\bar{v}$ and the mean rotational velocity $\bar{\omega}$
is very small.

Experiments in 3D already indicate that 
the characteristics of the motion are hardly influenced by material
properties \cite{agu95,bid}. Rolling steel, glass and plastic balls
down the plane gives nearly the same velocity, though for
plastic (which has the lowest $e_n$ and largest $\mu$) the B region of
the phase diagram is found to be somewhat extended. 
Our simulations in 2D indeed show that the mean velocity $\bar{v}$ is
nearly independent of both $e_n$ and $\mu$. Fig.~11 demonstrates this
for the case of a ball of size ratio $\Phi=2.25$. Fig.~11a shows the
velocities for varying normal coefficient of restitution $e_n$. It can
clearly be seen that $e_n$ influences the phase diagram, {\em
  i.e.}~the extension of region B, but has only a very small influence
on $\bar{v}$. Though $\theta_{\mbox{\tiny AB}}$ is hardly
affected by $e_n$ (except for very small $\Phi$), $\theta_{\mbox{\tiny
    BC}}$ moves to larger inclination angles 
$\theta$ with decreasing $e_n$. But $e_n$ influences $\bar{v}$
slightly in a direction that is contrary to what one would expect
intuitively. Increasing the dissipation leads to a
slight increase in the velocity. We have found no explanation for this
so far. We have, however, understood the relative insensitivity of
$\bar{v}$ to $e_n$. 

\begin{figure}[tbh]
\centerline{
\epsfysize 5 cm
\epsfbox {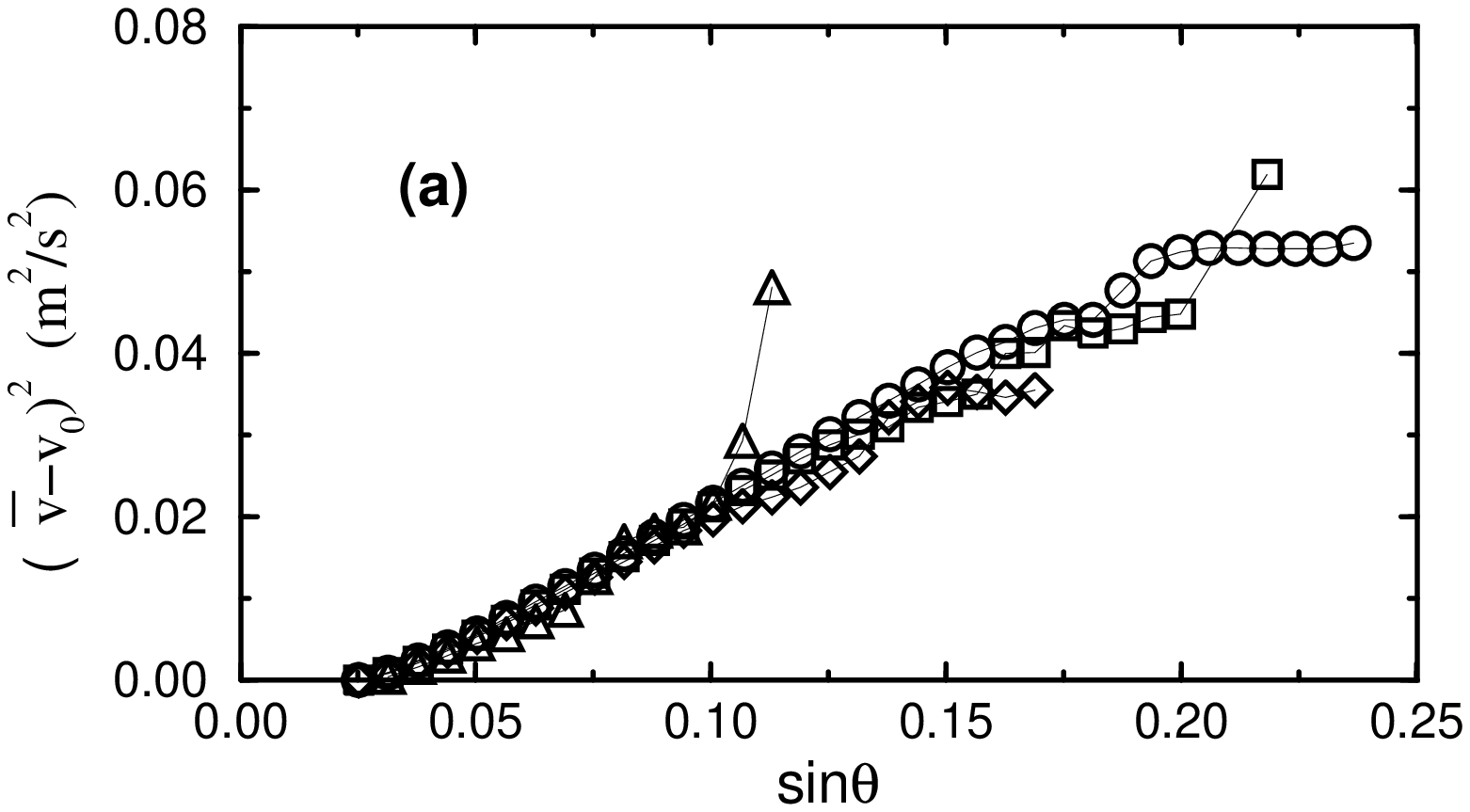}}
\centerline{
\epsfysize 5 cm
\epsfbox {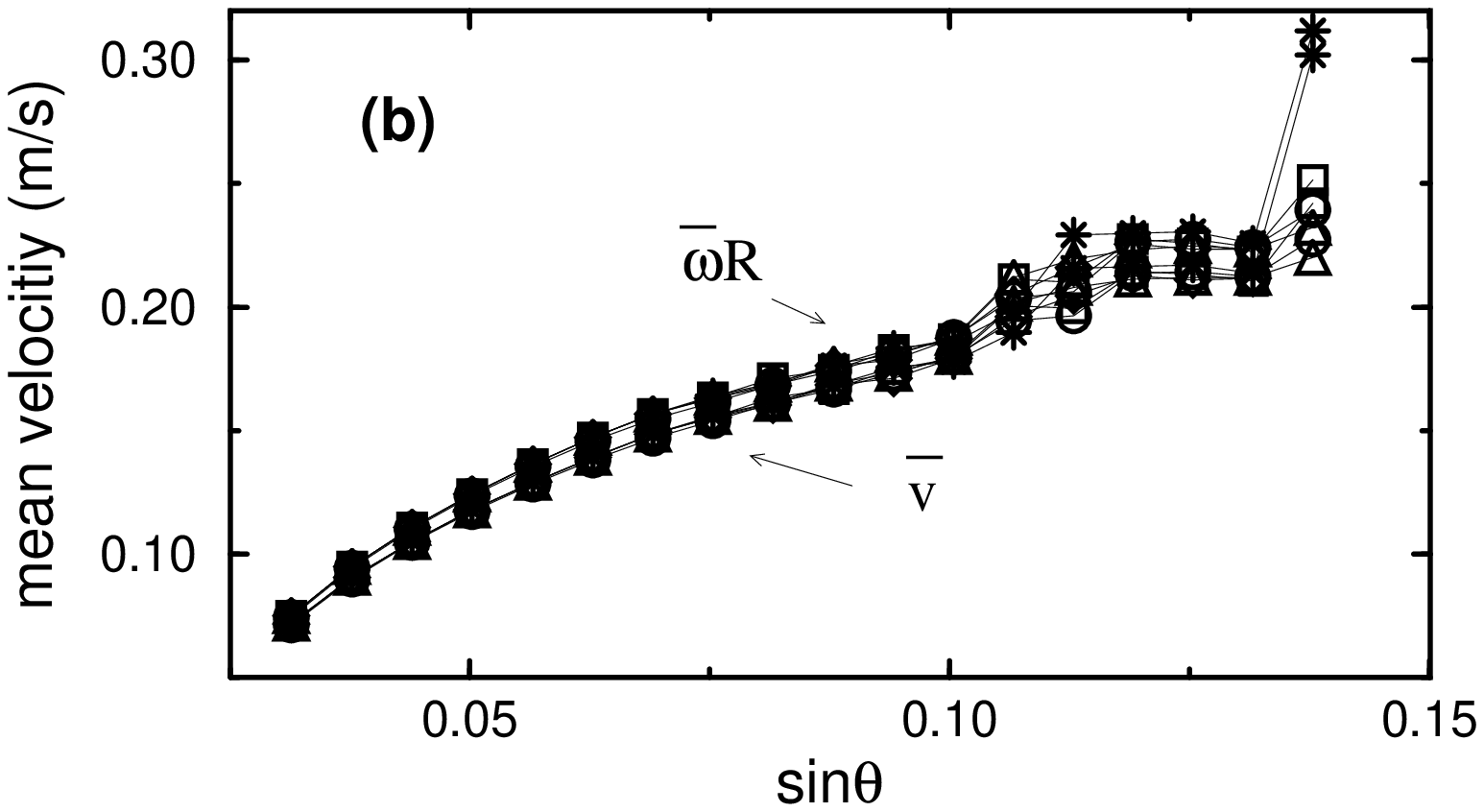}}
{\small FIG.~11. (a) Dependence of $\bar{v}$ on the dissipation for
  $\Phi=2.25$ and $\epsilon=0$: $e_n=$
  0.4($\bigcirc$), 0.5($\Box$), 0.6($\Diamond$),
  0.8($\bigtriangleup$).  (b) Dependence of
  $\bar{v}$ and $\bar{\omega}R$ on the friction coefficient $\mu$:
  $\mu=$ 0.1($\bigcirc$), 0.3($\Box$), 0.5($\Diamond$),
  0.7($\bigtriangleup$), 1.0($\star$). 
}
\end{figure}

The reason for the small influence of $e_n$ is essentially the
regularity of the motion. As shown in the previous subsection, the
ball moves over each line ball in a succession of bounces, in
the course of which it loses all or most of its relative normal
velocity with respect to this line ball. $e_n$ mainly determines how
many bounces are necessary to achieve this. As long as the moving ball
has only a negligible amount of normal velocity left as it reaches
$\gamma_{\mbox{\tiny max}}$, it seems to be unimportant for the
tangential velocity at this point how many impacts were needed to lose
the normal velocity the ball had at $-\gamma_{\mbox{\tiny max}}$. 

For the phase boundary BC, the value of $e_n$ is important. Since we
found that the motion of the ball starts to get unstable when the
motion is no longer regular, we expect the
destabilization to start when the normal velocity at
$-\gamma_{\mbox{\tiny max}}$ with respect to one ball cannot be lost
by the time $\gamma_{\mbox{\tiny max}}$ is reached. But $e_n$,
which determines the energy loss in a collision and thus the height
(and thereby the length) of the next jump, together with $\bar{v}$
determines how many jumps are made on one line ball. It thus gives an
upper limit to the maximum amount of energy that can be dissipated on
a single line ball. The smaller $e_n$, the more energy is dissipated
in each jump, and the more jumps are possible, as their length
decreases. Thus, $\theta_{\mbox{\tiny BC}}$ shifts to larger $\theta$
for a given $\Phi$ with decreasing $e_n$ and approaches the static
angle of stability, which in the 2D case is given by
$\gamma_{\mbox{\tiny max}}$.

We find that because the ball is allowed
to rotate, the value of the friction coefficient $\mu$ is even less
important for the behaviour of the ball than $e_n$ (see
Fig.~11b). $\mu$ influences 
neither $\bar{v}$ nor the mean rotational velocity $\bar{\omega}$
significantly. It also does not change the phase boundaries.
The reason for this is that for the range of $\theta$ considered here,
the ball can be expected to roll without slipping most of the time
independent of $\mu$ if our
implementation of the tangential force law correctly reproduces
Coulomb's law for sliding friction. Let us assume for a moment that
the ball, while moving down the line, is always in contact with the
balls on this line, which is only possible if $e_n=0$. It would then
roll without slipping if at all $\gamma$ the condition
$|F_s|\leq\mu|F_n|$ were fulfilled. For a rolling sphere under the
action of gravity \cite{rad95a},  
\beq 
|F_s| = \frac{2}{7} |\vec{g} \cdot \vec{s}|,
\eeq
so that the criterion for $\mu$ for the ball to roll without slipping
reads 
\beq
\mu > \frac{2}{7} |\tan(\theta+\gamma)|.
\label{slide}
\eeq
Though close to $\gamma_{\mbox{\tiny max}}$ this condition is usually
not fulfilled (the smaller $\Phi$, the larger the region of $\gamma$
for which slip can occur), it holds on the
largest part of a ball on the line even for small $\mu$. We would thus
assume the ball on the average to roll without slipping in the case of
$e_n = 0$ even for small values of the friction coefficient $\mu$. 

In our simulations, however, we used larger values for $e_n$, like
$e_n=0.7$ in 
Fig.~11b, so there the ball rather bounced than rolled down the line. 
But the distances the ball covers between bounces in the steady state
are very small. We find in our simulations that
the ball, which we launch onto the plane without rotational velocity,
soon picks up rotation during impacts, such that when the steady state
is reached, the rotational velocity has adjusted itself to a value
which on average leads to zero relative velocity of the surfaces of
the 
moving ball and the fixed balls (except close to $\gamma_{\mbox{\tiny
    max}}$, just as would be expected from eq.~(\ref{slide})).  In the
free flight between collisions, the ball picks 
up translational velocity while the rotational velocity remains
unchanged. The moving ball has thus gained excess shear velocity,
which is converted nearly completely into
rotation in the next impact, if it was small enough. This is usually
the case in the steady state, where distances covered between bounces
are small.

\section{A theoretical model for the mean velocity}

In this section, we derive $\bar{v}$ by a simple analytical
treatment. Our simulation results play an important role here, as they
show which simplifications can be introduced without losing essential
features of the motion.

From our results on the influence of the coefficient of restitution
$e_n$ on the motion of the ball, we can deduce a very accurate result
for $\bar{v}$ in the case of equally spaced balls on the line. Since
we find that $\bar{v}$ hardly depends on $e_n$ 
we investigate the limiting case $e_n=0$. As we also found that,
due to friction, the ball is able to sustain $v_s=0$ on average ({\em
  i.e.~}the ball would roll without slipping in the case $e_n=0$ for
large $\mu$), we make the following assumptions.

We assume that the moving ball is always in contact with the fixed
balls, {\em i.e.~}rolls down the line without slipping. Thus at all
times, 
\begin{equation}
|\vec{v}| = v_t = \vec{v} \cdot \vec{s},
\end{equation}
\begin{equation}
v_n=0,
\end{equation}  
\begin{equation}
\omega R = v_t.
\label{omega}
\end{equation}

From these conditions the equation of motion of the
ball on a single ball on the line from $-\gamma_{\mbox{\tiny max}}$ to
$\gamma_{\mbox{\tiny max}}$ can be derived. The kinetic energy of the
ball is 
\begin{equation}
E_{\mbox{\tiny kin}} = \frac{1}{2} m v_t^2 + \frac{1}{2} J \omega^2
\end{equation}
where $J$ denotes the moment of inertia of the moving particle, which
in the case of a sphere of radius $R$ takes the value
$J=\frac{2}{5}mR^2$. Using condition (\ref{omega}), we get
\begin{equation}
E_{\mbox{\tiny kin}} = \frac{7}{10}m v_t^2
\label{ekin}
\end{equation}
for a sphere. The potential energy depends on the location of the
moving ball on the line ball:
\begin{equation}
E_{\mbox{\tiny pot}} = m g (R + r) \cos(\gamma+\theta).
\end{equation}
Since $v_t=(R+r)\dot{\gamma}$, the energy balance reads (denoting the
energies at the start of the motion by $E_{\mbox{\tiny kin}}^0$ and
$E_{\mbox{\tiny pot}}^0$)
\begin{equation}
E_{\mbox{\tiny kin}}^0 - \frac{7}{10}m(R+r)^2 \dot{\gamma}^2
=  m g (R + r) \cos(\gamma+\theta) - E_{\mbox{\tiny pot}}^0.
\label{energies}
\end{equation}
Differentiating with respect to time yields the equation of motion for
$\gamma(t)$
\begin{equation}
\ddot{\gamma} = \frac{5}{7} \frac{g}{R+r} \sin(\gamma+\theta).
\label{motion}
\end{equation}
The completely inelastic collision at $-\gamma_{\mbox{\tiny max}}$
that occurs when the moving ball passes from one line ball to the next
defines the boundary conditions for the problem.
We denote the velocity of the ball at $\gamma_{\mbox{\tiny max}}$ by
$v_t(\gamma_{\mbox{\tiny max}})=v_i$. Since we assume
rolling without slipping, on this ball on the line we had
\begin{equation}
v_i = \omega_i R
\label{omegai}
\end{equation}
with $\omega_i = \omega(\gamma_{\mbox{\tiny max}})$.
In the next instant, the moving ball hits the next ball on the line at
$-\gamma_{\mbox{\tiny max}}$, with the tangential velocity 
$v_t(-\gamma_{\mbox{\tiny max}})$, which with respect to this ball is
(due to the change in geometry) 
\begin{equation}
v_t(-\gamma_{\mbox{\tiny max}}) =  v_i\cos(2\gamma{\mbox{\tiny max}}). 
\end{equation}
In the collision that is now about to take place, the normal component
\begin{equation}
v_n(-\gamma_{\mbox{\tiny max}}) =  v_i\sin(2\gamma{\mbox{\tiny max}})
\end{equation}
is reduced to zero, but this is not the whole effect of the collision.
Since $v_t$ dropped {\em before} the impact due to the change in
geometry, but $\omega$ was left unaffected, at $-\gamma_{\mbox{\tiny
    max}}$, there is excess rotational velocity and thus excess shear
velocity $v_s$. If the frictional force is high
enough (which we assume in this treatment), then the shear velocity at
the point of contact should again be reduced to zero during the
impact, such that tangential and rotational velocity {\em after} the
impact at $-\gamma_{\mbox{\tiny max}}$, which we will denote by $v_f$
and $\omega_f$ respectively, fulfill the condition  
\begin{equation}
v_f = \omega_f R.
\label{omegaf}
\end{equation}
During the impact at $-\gamma_{\mbox{\tiny max}}$ the rotational and
translational velocities thus adjust themselves, with only
negligible  energy
loss (friction here mainly helps to distribute the excess rotational
velocity to the translational degree of freedom, but, as the ball can
rotate freely, dissipates only very little energy during the
adjustment). Before the adjustment of 
rotational and translational velocities, the kinetic energy of the
system was
\begin{equation}
E_{\mbox{\tiny kin}}= \frac{1}{2} m v_i^2 \cos^2(2\gamma_{\mbox{\tiny max}}) +
\frac{1}{2} J \omega_i^2
\end{equation}
With eqs.~(\ref{ekin}) and (\ref{omegaf}) we thus obtain  
from the energy balance the condition for $v_f$ 
\begin{equation} 
\frac{1}{2}v_i^2 \cos^2(2\gamma_{\mbox{\tiny max}})+\frac{1}{2}J\omega_i^2 =
\frac{7}{10}mv_f^2. 
\end{equation}
Substituting $\omega_i$ according to eq.~(\ref{omegai}), we get 
\begin{equation}
v_f=v_i\sqrt{\frac{1}{7}( 5 \cos^2(2\gamma_{\mbox{\tiny max}}) + 2)}
\label{boundary}
\end{equation}
for the tangential velocity after the collision at
$-\gamma_{\mbox{\tiny max}}$. Since $v_t=(R+r)\dot{\gamma}$, this
provides the boundary condition for eq. (\ref{motion}). 

We obtain $\bar{v}$ numerically by starting at $-\gamma_{\mbox{\tiny
    max}}$ with an arbitrary and very high starting velocity
$\dot{\gamma}_0$, letting the system evolve according to
eq.~(\ref{motion}). Whenever $-\gamma_{\mbox{\tiny max}}$ is reached,
eq.~(\ref{boundary}) is applied, and we start again at
$\gamma_{\mbox{\tiny max}}$. We continue this process until either
the ball rolls back (which we consider as trapping) or until it
reaches a steady state. The mean velocity in this steady state is
plotted in Fig.~12, in comparison with simulation data for various
coefficients of restitution. Clearly, the simulation for small
$e_n$ is closest to the theoretical curve (which assumes $e_n=0$), but
also for higher $e_n$ the approximation is still good.
    
For the case $\epsilon>0$ eq.~(\ref{motion}) and (\ref{boundary})
yield equally good results as in the case $\epsilon=0$ presented in
Fig.~12. In the case of a disordered line, an estimation of $\bar{v}$
with the value of $\gamma_{\mbox{\tiny max}}$ chosen according to the
mean value of $\epsilon$ fits the simulation results equally well.

\begin{figure}[tbh]
\centerline{
\epsfysize 5 cm
\epsfbox {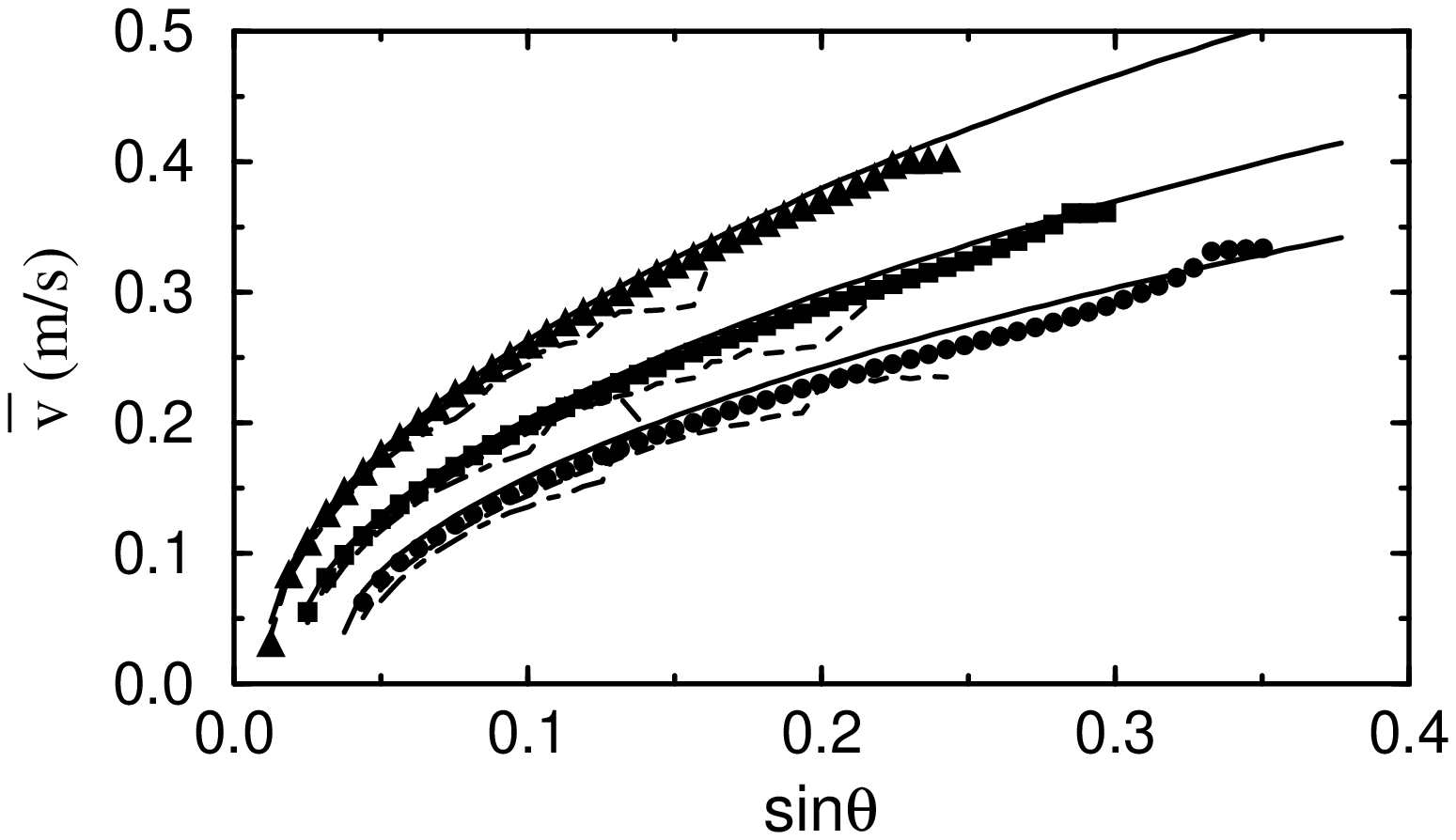}}
{\small FIG.~12. Theoretical prediction of $\bar{v}$ (solid lines) for
  $\epsilon = 0$ 
  under assumption of $e_n=0$ and simulation data for $e_n=0.1$ and
  size ratios $\Phi=$1.75 (circles), 2.25 (squares), 3.0
  (triangles). The dashed and dot-dashed lines respectively
  also show the corresponding simulation data for $e_n=0.5$ and
  $0.7.$} 
\end{figure}

Another result that can be obtained from this theoretical treatment is
the phase boundary $\theta_{\mbox{\tiny AB}}$. To this end, we make
use of eq.~(\ref{energies}). By setting  
\begin{equation}
E_{\mbox{\tiny kin}}^0=
\frac{7}{10}m(R+r)^2\dot{\gamma}^2(-\gamma_{\mbox{\tiny max}}) 
\end{equation}
and 
\begin{equation}
E_{\mbox{\tiny pot}}^0= mg(R+r)\cos(\theta-\gamma_{\mbox{\tiny max}}),
\end{equation}
the values at the begin of the motion over one ball in the line,
and by using the fact that in the steady state
\begin{equation}
\dot{\gamma}(-\gamma_{\mbox{\tiny
    max}})=e_t\dot{\gamma}(\gamma_{\mbox{\tiny max}}) 
\end{equation}
with 
\begin{equation}
e_t= \sqrt{\frac{1}{7}( 5 \cos^2(2\gamma_{\mbox{\tiny max}}) + 2)},
\end{equation}
we obtain from eq.~(\ref{energies}) the starting velocity
$\dot{\gamma}(-\gamma_{\mbox{\tiny max}})$ in the steady state
\begin{equation}
\dot{\gamma}^2(-\gamma_{\mbox{\tiny
    max}}) =
\frac{20}{7}\frac{g}{R+r}\frac{e_t^2}{1-e_t^2}\sin\gamma_{\mbox{\tiny
    max}}\sin\theta.
\label{steady}
\end{equation}
We now assume that the phase boundary $\theta_{\mbox{\tiny AB}}$ is
reached when the moving ball arrives at the angle $\gamma=-\theta$ with
zero velocity, since from there it can roll down simply by the action
of gravity, even with zero starting velocity at this point. For smaller
inclination angles, the ball would roll back before reaching this
point and thus stop, for larger inclination angles it would pass this
point with some velocity and move on. From eq.~(\ref{energies}) we
obtain 
\begin{equation}
\dot{\gamma}^2(-\gamma_{\mbox{\tiny
    max}}) =
\frac{10}{7}\frac{g}{R+r}\left(1-\cos(\theta_{\mbox{\tiny AB}}-\gamma_{\mbox{\tiny
      max}})\right)
\label{stop}
\end{equation}
by setting $\gamma=-\theta_{\mbox{\tiny AB}}$ and
$\dot{\gamma}(-\theta_{\mbox{\tiny AB}}) = 0$.
If the ball is in region B, the steady state condition (\ref{steady})
has to be fulfilled as well, so that from eq.~(\ref{steady}) and
eq.~(\ref{stop}) we get an equation for $\theta_{\mbox{\tiny AB}}$
\begin{equation}
\frac{2e_t^2}{1-e_t^2}\sin\gamma_{\mbox{\tiny
    max}}\sin\theta_{\mbox{\tiny AB}} = 1-\cos(\theta_{\mbox{\tiny
    AB}}-\gamma_{\mbox{\tiny max}}),
\end{equation}
which finally yields
\begin{eqnarray}
\sin\theta_{\mbox{\tiny AB}} & = & \frac{\sin\gamma_{\mbox{\tiny
      max}}}{\left(\frac{1+e_t^2}{1-e_t^2}\right)^2\sin^2\gamma_{\mbox{\tiny max}}+\cos^2\gamma_{\mbox{\tiny max}}}\cdot\\\label{phase}
& & \left(\frac{1+e_t^2}{1-e_t^2}-\frac{2e_t}{1-e_t^2} \cos\gamma_{\mbox{\tiny max}}\right).\nonumber
\end{eqnarray}

Fig.~13 shows a comparison of this result to simulation
data. Obviously, our theoretical result approximates the simulation
results best for larger $\Phi$. For $\Phi$ close to 1 the deviations
get quite large, since here the assumption that the ball loses its
normal velocity in a single impact is not fulfilled any more as well
as before even for $e_n=0.1$. In any case, eq.~(\ref{phase}) provides
a lower limit for the value of $\theta_{\mbox{\tiny AB}}$.

\begin{figure}[tbh]
\centerline{
\epsfysize 5 cm
\epsfbox
{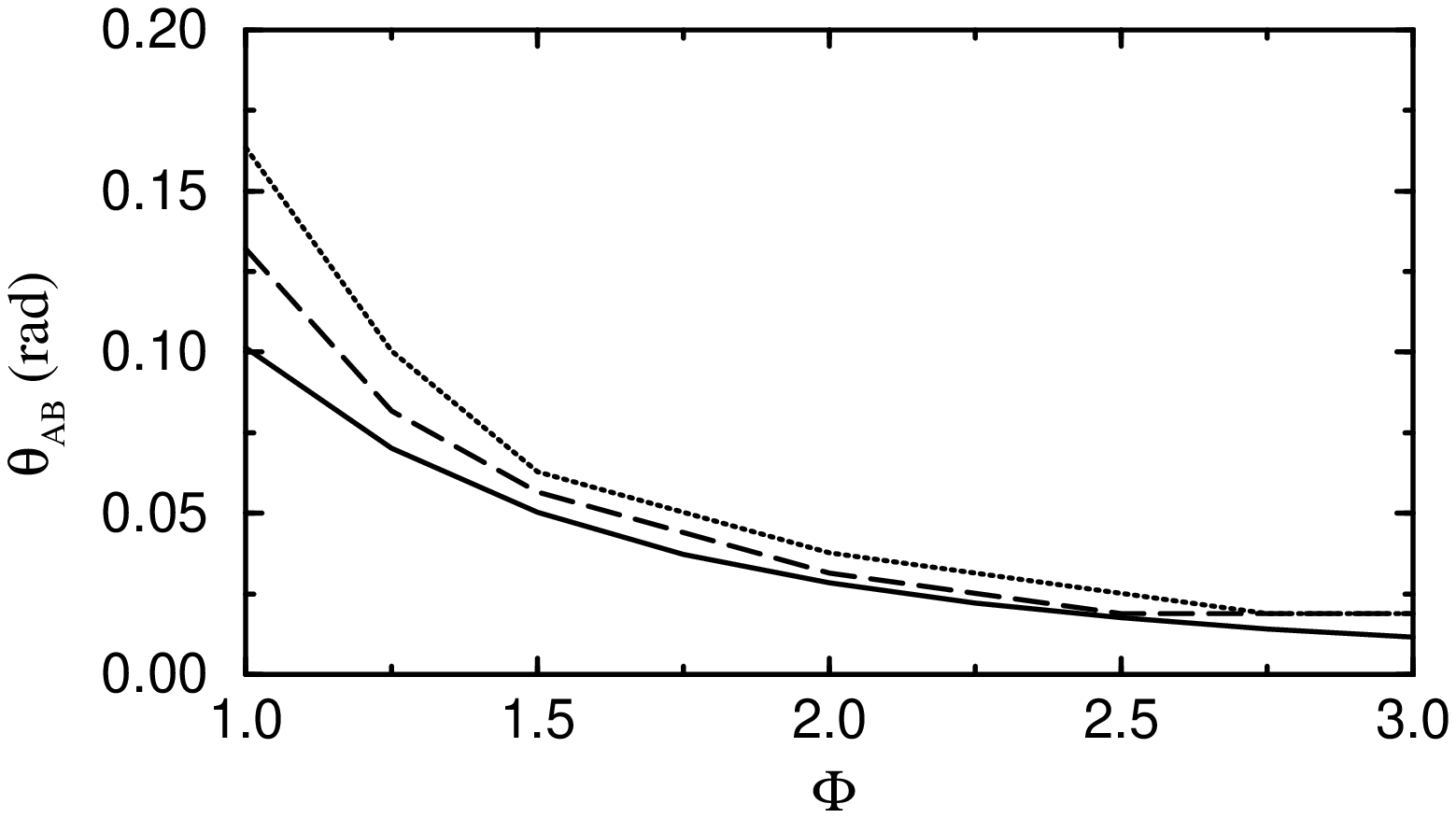}} 
{\small FIG.~13.  The phase boundary $\theta_{\mbox{\tiny AB}}$ for
  $e_n=0.7$ (dotted line), $e_n = 0.1$ (dashed line) and the
  theoretical result from eq.~(\ref{phase}) for $e_n = 0$ (solid
  line). } 
\end{figure}

After the completion of this work, we became aware of work by
Ancey {\em et al.~} \cite{anc96}
paralleling our theoretical treatment. They, however, neglect the
effect of the rotational velocity in the impact at
$-\gamma_{\mbox{\tiny max}}$, though rotation is explicitly included
in their equation of motion. They thus have to introduce a fitting
parameter to match their curves to experimental data. This is not
necessary if the influence of rotation is included in the boundary
conditions of the problem, as we have shown. We emphasize that
eqs. (\ref{motion}) and (\ref{boundary}) hold for the {\em
  instantaneous} velocity only in the case of
vanishing normal restitution and perfectly rough surfaces, conditions
which are not likely to be fulfilled by any commonly used
material. However, as we have shown, the {\em mean} velocity of the 
ball is not influenced by the coefficients of restitution and friction
and thus is correctly described. 

One more result can be deduced directly from our theoretical
treatment. Since $\dot{\gamma}\sim(r(\Phi+1))^{-1/2}$, and
$v_t=(\Phi+1)r\dot{\gamma}$, we expect $\bar{v}$ scale with $\sqrt{r}$,
{\em i.e.~}to depend on the {\em absolute} size of the balls. Note
that it does not, however, simply scale with $\sqrt{\Phi+1}$ as well,
as  $\Phi$ also enters into the boundary conditions and thus changes
$\dot{\gamma}$. But,
keeping the geometry of the system ({\em i.e.~}$\Phi$ and $\epsilon$)
constant, we expect a scaling of $\bar{v}$ with $\sqrt{r}$ in our
simulations. 

This scaling with $r$ is observed in our simulations (see Fig.~14). In
2D experiments, it had been observed in an 
experimental setup restricted to $\Phi=1$ \cite{jan92}, here we find
it as well if we keep $\Phi$ fixed at an arbitrary value. This scaling
seems to be quite universal, since it was also found in 3D experiments
\cite{lilpriv} and in the 2D stochastic model \cite{ggb}. 

\begin{figure}[tbh]
\centerline{
\epsfysize 5 cm
\epsfbox {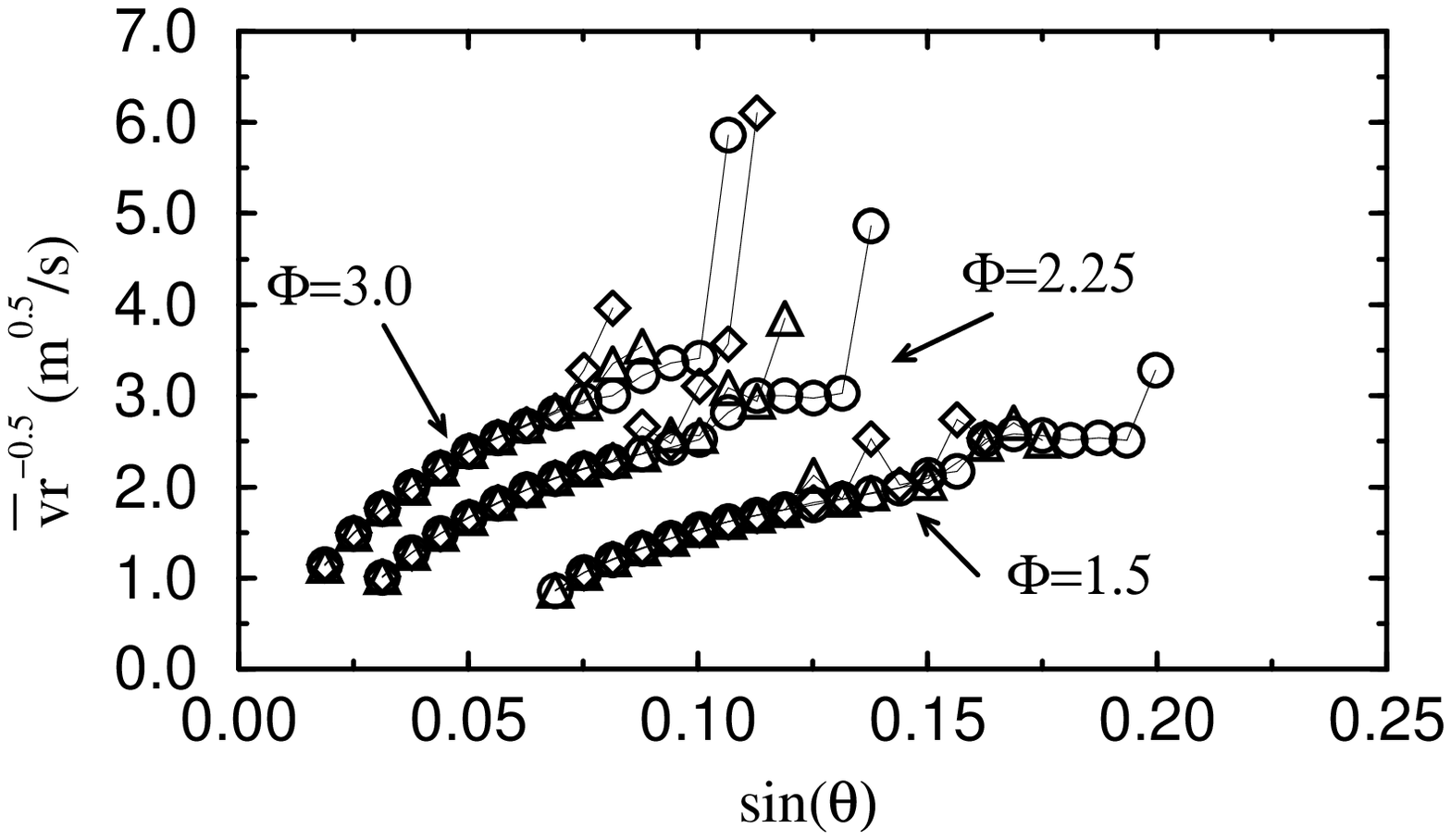}}
{\small FIG.~14. Scaling of $\bar{v}$ with $r$ keeping $\Phi$
  constant. $r=$0.5 mm ($\diamond$), 1.25 mm ($\bigtriangleup$), 5 mm
  ($\bigcirc$). }
\end{figure}
 
\section{Summary and discussion}

We have discussed the mechanism stabilizing the motion of a ball on an
inclined line consisting of equally sized balls.
Two important results emerged from our simulations. Firstly, we found
that the motion of the ball in the steady state is very regular
and consists of a series of small bounces on {\em
  each} ball on the line, contrary to what so far has been assumed
\cite{ris94b,ggb}. In the course of these bounces the moving ball
loses all relative normal 
velocity with respect to this ball. This dissipation mechanism holds
both for lines 
with equally spaced balls and for lines with a random spacing of
balls. Furthermore, the location of these bounces is the same on each
ball, so that a random spacing of the balls on the line only smears
out these locations a bit, but does not alter the motion
significantly. Thus, the velocity of the ball in the case of a random
spacing of balls on the line can be 
approximated by the motion of the ball on an ordered line with
appropriate spacing. The main reason for this regularity of the motion
is that the moving ball has to climb over the top of every ball
forming the line by a few bounces in the steady state. Since disorder 
only slightly changes the height to be climbed, it has only a small
influence on the motion. 

Clearly, this mechanism for keeping a constant velocity cannot hold in
3D, where there is a number of ways for the ball to choose 
to go down the plane. The 3D case should be ruled by a
competition between stability properties of the plane and the
preferential direction given by the plane inclination and inertia of
the moving ball. One might argue that introducing stronger disorder in
the 2D case, for example by using polydisperse balls on the line,
would model a longitudinal section of the plane much closer. But 
this will not eliminate
the strong dimensional difference between the 2D and 3D cases. In 3D,
the moving ball, hitting a ball on the plane a little on the side
(which would be one of the smaller balls in the 2D section), would be
deflected towards the side, which in 2D is impossible. 
In the constant velocity regime in 2D on a line with size
polydispersity, the moving ball will again lock into some kind of
quasiperiodic behaviour for the same reasons as explained above. The
problem of having to overcome some maximum threshold (to
get out of 
the deepest valley between two balls on the line) to keep moving is
even stronger, contrary to the 3D case. There a ball, having suffered
a 
large impact which brakes it a lot, might still find some way around
the bump that braked it and accelerate until the next  
impact via some statically unstable path. Thus, the motion in 3D is
expected to be far more irregular than can be modeled in 2D. 
Future simulations in 3D will have to show where these
differences originate and what the stabilizing mechanism consists of
in this case. A stochastic model for the 2D case \cite{ggb,proc}
nevertheless showed a viscous friction force, {\em i.e.~}a linear
dependence of $\bar{v}$ on $\sin\theta$. This might be due to the fact
that the randomness introduced there in the choice of the next impact
captures the randomness of the ``real'' 2-dimensional
plane. However, this question can only be answered ultimately by 3D
simulations.  

The second important result is that material properties like the
normal coefficient of restitution $e_n$ and the Coulomb friction
coefficient $\mu$ hardly influence the mean velocity $\bar{v}$ in the
steady state. From this result and the knowledge of the mechanism
stabilizing the motion of the ball 
in 2D we were able to predict $\bar{v}$ theoretically under the
assumption of completely 
inelastic collisions. Only geometrical considerations sufficed for
this treatment and gave a very good approximation to our simulation
results. The results in this extreme case enabled us to derive a lower
limit for the phase boundary $\theta_{\mbox{\tiny AB}}$, which for
large $\Phi$ gives a good approximation of $\theta_{\mbox{\tiny AB}}$ 
regardless of the value of the coefficient of restitution. This result
should be relevant to the problem of segregation in the flow on inclined
planes or in rotating drums. 

Though we have made significant progress in understanding the
stabilizing mechanism by which the ball 
keeps its constant velocity in 2D, a few open questions remain. 
Due to the strong nonlinearity of the problem (computation of
successive impact angles, velocities etc.~from the equations of motion
would require the solution of a fourth order polynomial), a direct
iteration of the equations of motions as for $e_n=0$ is very
cumbersome, though in principle possible, in
the case of non-vanishing $e_n$. Since the phase boundary
$\theta_{\mbox{\tiny BC}}$, which separates steady motion of the ball
from the accelerating regime, depends on $e_n$, it cannot be derived
directly from our theoretical treatment. The most obvious
simplification one might introduce, namely ignoring the structure of
the plane in the computation of the next impact, is out of the
question, since we have found this structure to be essential for the
stabilizing mechanism. The 
problem seems to be related to the classical problem of a ball
bouncing on a vibrating plate \cite{lie72}, which lately has been
discussed for the case of a partially inelastic ball \cite{luc93}. In
the case of finite restitution, it was found that neglecting the
motion of the vibrating plate can lead to erraneous results, like the
observation of ``chaos'' in a region of phase space where a
more exact treatment shows the existence of eventually periodic
orbits. Unfortunately, even this simple one-dimensional problem can
only be solved by approximations in the extreme cases $e_n\ll 1$ and
$e_n\rightarrow 1$, so it is by no means obvious how to relate these
results to our problem, since for these extreme cases, we already
have solved the problem. (The case $e_n\rightarrow 1$ is trivial,
since there region B of the phase diagram vanishes.)

So for the 2D case, the remaining open questions are the
following. So far, we do not understand well how the
ball gets braked down to its 
stable velocity when it is launched on the plane with a much higher
initial velocity, or why it accelerates as soon
as it bounces only approximately once per ball on the line. In
addition, 
the questions of whether in the bouncing region a steady state is
reached as 
well, as has been suggested before \cite{ris94b}, and what the
characteristics of this steady state are, are of interest. Does the
ball really move in a ``chaotic'' way in the bouncing regime,
as is suspected \cite{ris94b}, or might an eventually periodic motion
exist as in the case of the ball on the vibrating plate, which due to 
the long transient so far could not be observed?

Although many of the interesting features of granular flow come about
by the {\em collective} behaviour and the interaction of many
particles, we have shown that even the elementary processes involving
only a single particle moving in a dissipative, but fixed environment,
can
offer quite a few new insights and open up new questions. One of these
questions, which is also of relevance to granular flows in general, is
that of dimensionality. It has always been implicitly assumed that
2-dimensional model systems are an adequate tool to study granular
flows 
in general, {\em i.e.~}that the qualitative behaviour will not be 
different in 3D. Our results show that even though this is probably
true for behaviour {\em in} the bulk, where grain motion is very
confined, care has to be taken in relating the behaviour on free
surfaces in 2D to that in 3D.

\section*{Acknowledgments}

We wish to thank D. Bideau, I. Ippolito, L. Samson and J. Sch\"afer
for very valuable discussions. This work was supported in part by the
Groupement de Recherche CNRS \ltue Physique des Milieux
H\'et\'erog\`enes Complexes\rtue~and by the HCM European Network \ltue
Cooperative Structures in Complex Media\rtue.

\end{multicols} 


\begin{thebibliography}{99}

\bibitem{dra90}
  T. G. Drake,
  J.~Geophysical Reserch {\bf 95}, 8681 (1990)
 
\bibitem{sav89}
  S. B. Savage, K. Hutter,
  J. Fluid Mech.~{\bf 199}, 177 (1989)
 
\bibitem{sav93}
  S. B. Savage,
  in: ``Disorder and Granular Media", D. Bideau and A. Hansen
  (eds.), Elsevier, Amsterdam, 1993
 
\bibitem{can95a}
  F. Cantelaube, D. Bideau,
  Europhys.~Lett.~{\bf 30}, 133 (1995)
 
\bibitem{cle95}
  E. Cl\'{e}ment, J. Rajchenbach, J. Duran, 
  Europhys. Lett.~{\bf 30}, 7 (1995)


\bibitem{zik94}
  O. Zik, D. Levine, S. G. Lipson, S. Strikman, J. Stavans,
  Phys.~Rev.~Lett.~{\bf 73}, 5 (1994)

\bibitem{bau95}
  G. Baumann, I. M. J\'anosi, D. E. Wolf,
  Phys.~Rev.~E {\bf 51}, 1879 (1995)

\bibitem{cam90}
  C. S. Campbell,
  Ann.~Rev.~Fluid Mech.~{\bf 22}, 57 (1990)

\bibitem{fau95}
  M. Caponeri, S. Douady, S. Fauve, C. Laroche, 
   in: {\em Mobile Particulate Systems},
  ed.~E. Guazzelli and L. Oger, Kluwer, Dordrecht, 1995
 
\bibitem{pou95b}
  O. Pouliquen, N. Renaut, 
  J. Phys.~II (France) {\bf 6}, 923 (1996)

\bibitem{rig94a}
  F.-X. Riguidel, R. Jullien, G. H. Ristow, A. Hansen, D. Bideau,
  J.~Phys.~I (France) {\bf 4}, 261 (1994)
 
\bibitem{rig94b}
  F.-X. Riguidel, A. Hansen, D. Bideau,
  Europhys. Lett.~{\bf 28}, 13 (1994)
 
\bibitem{ris94b}
  G. H. Ristow, F.-X. Riguidel, D. Bideau,
  J.~Phys.~I (France) {\bf 4}, 1161 (1994)
 
\bibitem{agu95}
  A. Aguirre, I. Ippolito, A. Calvo, C. Henrique, D. Bideau,
  unpublished 

\bibitem{jan92}
  C. D. Jan, H. W. Shen, C. H. Ling, C. I. Chen, Proc. of the 9th
  Conf. of Eng. Mech., p. 769 (1992)

\bibitem{rigdiss}
  F.-X. Riguidel, PhD thesis, Univ. Rennes, France (1994)

\bibitem{bag54}
  R. A. Bagnold,
  Proc.~Roy.~Soc.~A {\bf 225}, 49 (1954)

\bibitem{ggb}
  G. G. Batrouni, S. Dippel, L. Samson, 
  Phys.~Rev.~E {\bf 53}, 6496 (1996) 

\bibitem{all87}
  M. P. Allen, D. J. Tildesley,
  {\em Computer Simulation of Liquids},
  Clarendon Press, Oxford, 1987

\bibitem{cun79}
  P. A. Cundall, O. D. L. Strack,
  G\'eotechnique {\bf 29}, 47 (1979)

\bibitem{schae96}
  J. Sch\"afer, S. Dippel, D. E. Wolf,
  J. Phys.~I (France) {\bf 6}, 5 (1996)

\bibitem{foe94}
  S. F. Foerster, M. Y. Louge, H. Chang, K. Allia,
  Phys.~Fluids {\bf 6}, 1108 (1994)

\bibitem{rad96}
  F. Radjai, J. Sch\"afer, S. Dippel, D. E. Wolf, HLRZ-Preprint
  30/96 
  
 
\bibitem{proc}
  S. Dippel, L. Samson, G. G. Batrouni, Proceedings of ``HLRZ
  Workshop on Traffic and Granular Flow", World Scientific, Singapore
  (1996)

\bibitem{schae95}
  J. Sch\"afer, D. E. Wolf,
  Phys.~Rev.~E {\bf 51}, 6154 (1995)

\bibitem{mcn92}
  S. McNamara, W. R. Young, Phys.~Fluids {\bf A4}, 496 (1992)

\bibitem{bid}
  D. Bideau, I. Ippolito, L. Samson, G. G. Batrouni, S. Dippel,
  A. Aguirre, A. Calvo, C. Henrique, Proceedings of ``HLRZ Workshop
  on Traffic and Granular Flow", World Scientific, Singapore (1996)

\bibitem{rad95a}
  F. Radjai, S. Roux,
  Phys.~Rev.~E {\bf 51}, 6177 (1995)

\bibitem{anc96}
  C. Ancey, P. Evesque, P. Coussot,
  J.~Phys.~I (France) {\bf 6}, 725 (1996)
 
\bibitem{lilpriv}
  L. Samson, I. Ippolito, private communication

\bibitem{lie72}
  M. A. Liebermann, A. J. Lichtenberg, Phys.~Rev.~A{\bf 5}, 1852
  (1972) 

\bibitem{luc93}
  J. M. Luck, A. Mehta,
  Phys.~Rev.~E {\bf 48}, 3988 (1993)
 


\end{thebibliography}
\end{document}